\documentclass[11pt,a4paper]{article}
\usepackage[latin1]{inputenc}
\usepackage{bm}
\usepackage{microtype}
\usepackage{setspace}
\usepackage{mathtools}
\usepackage{extpfeil}
\usepackage{graphicx}
\usepackage[usenames,dvipsnames]{xcolor}
\usepackage{verbatim}
\usepackage{caption}
\usepackage{subcaption}
\usepackage{epstopdf}
\usepackage[margin=1in,footskip=0.25in]{geometry}
\usepackage[numbers, sort&compress]{natbib}
\usepackage{appendix}

\usepackage{hyperref}
\hypersetup{
	colorlinks=true,
	linkcolor=red,
	citecolor=NavyBlue
}

\newcommand{\er}{\bold{e}_r}

\newcommand{\tr}{\tilde{r}}
\newcommand{\tz}{\tilde{z}}
\newcommand{\tD}{\tilde{D}}
\newcommand{\ttime}{\tilde{t}}
\newcommand{\tu}{\tilde{u}}
\newcommand{\tw}{\tilde{w}}
\newcommand{\tJ}{\tilde{J}}
\newcommand{\tsigma}{\tilde{\sigma}}
\newcommand{\tmu}{\tilde{\mu}}

\begin{document}

\title{Rethinking battery degradation in presence of surface effects: mechanical versus electrochemical peformance mediated by charging condition}

\author{Amrita Sengupta, Jeevanjyoti Chakraborty}

\maketitle

%
%


\begin{abstract}
Surface stresses, in nano-sized anode particles undergoing chemomechanical interactions, have a relaxing effect on the diffusion-induced stresses thus improving the mechanical endurance of the particles, whereas, the compressive effect of surface stresses degrades the electrochemical performance. However, this straightforward prediction of an improved mechanical performance is challenged in this work. Silicon nanowires undergoing huge volumetric changes during lithiation, may undergo significant axial length-increase, which serves as an important criterion in determining the mechanical performance of SiNWs. Interestingly, surface stresses tend to reduce the length-increase under potentiostatic charging condition, but under galvanostatic charging, the length-increase gets enhanced, thus degrading the mechanical performance of the SiNWs. To further make the study more inclusive, the nanowire is modelled with a constraining material at its core, and a competitive analysis is presented for the overall performance of the anode particles under the combined effects of surface stresses and constraining material. The mathematical model is based on large deformation theory, considering two-way coupling of diffusion-induced stresses and stress-enhanced diffusion. It is hoped that this study will provide a fresh perspective in designing next-generation lithium-ion battery particles. 
\end{abstract} 

\clearpage


\section{Introduction}

The development of commercially-usable lithium-ion batteries (LIBs)\cite{mizushima1981lixcoo2,whittingham1979intercalation,yoshino2014development} led to a sudden upsurge in their demand over the last few decades. Due to its large-scale applications ranging from electronic gadgets to aircrafts, the focus has now shifted towards improving the storage capacity and energy density of LIBs. In order to fulfill the growing demand for high storage capacity, high energy and power densities, silicon (Si) as an anode material, with a theoretical capacity of 4200 mAh g$^{-1}$, is the best replacement for traditionally-used graphite (having a theoretical specific capacity of 372 mAh g$^{-1}$) \cite{huggins1999lithium,wu2018modeling,ozanam2016silicon}. The drawback associated with the usage of Si as anode is the large expansion in volume (upto 310 $\%$) of the anode particles upon insertion of lithium, and back to original volume upon dis-insertion \cite{beaulieu2001colossal, wang1999tem,kasavajjula2007nano,ko2015challenges}. This cyclic change in volume during each charging-discharging cycle leads to structural degradation\cite{tarascon2001issues,muller2018quantification,ebner2013visualization,kabir2017degradation,reniers2019review,liu2020safety,li2020comprehensive} of the anode particles, eventually lowering their electrochemical performance\cite{srivastav2017modelling,xu2015electrode}. The structural degradation could be in the form of buckling instabilities \cite{chakraborty2015influence,2016Bailinbuckling,2018BailinZheng,zhang2018lithiation,zhang85buckling}, fracture\cite{miehe2016phase, xu2019analytical,perassi2019capacity,swaminathan2007electrochemomechanical,swaminathan2007electrochemomechanical2,iqbal2020chemo, ram2020JAP}, altered material properties of the anode particles \cite{pan2020effect}, crack initiation \cite{liu2020cracks}, and a less-studied issue of length-increase in cylindrical particles\cite{chakraborty2015combining}. To mitigate these problems, researchers have come up with the solution of using nano-sized silicon anode particles \cite{arico2005nanostructured,lewis2007situ,chan2008high,graetz2003highly,xie2015phase,lai2019silicon,srijan2018JAP,nasr2019surface}. \\

With advancement in computational and manufacturing techniques, the research on different kinds of nano-structured anode particles have gained momentum, for example solid and hollow spherical particles, nanowires, and nanotubes \cite{chan2011high,wu2012designing,zhao2019review,liu2020computational,li2020real,heubner2020electrochemical,bunjaku2019structural,hu2020investigating,xu2019heterogeneous}. Nanostructured Si particles forming the anode in LIBs provide enough empty spaces around the particles to accommodate the large volume expansion associated with lithiation. Further, the time required for Li atoms to reach the core of the Si nanoparticles are less than that for larger particles. Thus, using nanoparticles, the time taken to reach full lithiation decreases. It is important to note here that, since the Li atom takes very less time to reach the Si core, the diffusion-induced stresses (DISs) generated within a Si particle (due to concentration-inhomogeneity of Li) is significantly low for nanoparticles. The lowering of DISs eventually lowers the chances of failure through fracture, thus increasing the structural integrity of the particles. Beside design perspectives, nano-wires and nano-tubes have enhanced electrochemical performance due to their robust electrical contact with the metallic current collector which enable each particle to contribute to the capacity \cite{chan2011high,wu2012designing}. A number of studies indicate the prominence of surface stress in nano-particles\cite{di2020shuttleworth} as the prime reason behind their structural durability \cite{cheng2008influence,deshpande2010modeling,hao2012diffusion,ASJC2019surfacestress,gao2015chemo,dingreville2008interfacial,schulman2018surface,jia2020coupling}. Furthermore, the effect of surface stress is dependent on the particle size (with decrease in particle size surface stress effect increases\cite{sharma2003effect,cammarata1994surface}), as well as the charging rate (with increase in charging rate, surface stress effect increases) \cite{ASJC2019surfacestress}. Hence, in order to maximize the mechanical endurance, we might decrease the particle size and increase the charging/discharging rate as much as possible but, it comes with a constraint in terms of the electrochemical performance of the electrode, or the battery as a whole. It has been observed for spherical Si anode particles undergoing small deformation, that the presence of surface stress degrades the electrochemical performance of the battery, and with increase in surface stress, the degradation increases \cite{lu2018reaction}. The suppressing or compressive effect due to surface stress at the surface of the particle, inhibits lithium uptake, thus lowering the charge capacity of the battery \cite{bucci2017effect}. Therefore, the surface stresses associated with nanoparticles give rise to two competitive phenomena where, on one hand, the particles become mechanically robust with decreasing tensile stresses, and on the other hand, the electrochemical performance is affected negatively. Strikingly, a decreasing tensile stress alone does not ensure an upgraded mechanical performance. Although it does ascertain a reduction in fracture and crack initiation, but another crucial mechanical phenomenon in case of axially unconstrained Si nanowires/nanotubes is the length-increase of the particles because of volumetric expansion during lithiation \cite{chakraborty2015combining}. The length-increase, along with the DISs, together will be considered as the mechanical performance of the electrode particles in our study. The entire analysis is performed for two different charging conditions: first, galvanostatic case where the charging/discharging rate is kept constant, and second, potentiostatic case where the interfacial voltage drop at the electrode-electrolyte interface is kept constant. The effect of surface stresses on the length-increase for different charging conditions has been analyzed in detail over very interesting and competitive results. To make our study comprehensive, we introduce a constraining material at the inner core of the Si nanotube, the properties of which modulates the length-increase of the particle \cite{chakraborty2015combining} which is already under the influence of surface stresses at the outer periphery.\\

The primary contribution of this work is an analysis of how charging condition plays a vital role in the surface-stress-induced competition between mechanical and electrochemical performances associated with reduced particle size in nano-structured cylindrical Si anode particles in a coupled, electro-chemo-mechanical environment, undergoing large deformation during each charging/discharging cycle due to insertion/dis-insertion of Li atoms. While developing the model for the Si nano-particle (owing to its huge volumetric expansion and contraction), we consider the modified surface stress formulation \cite{ASJC2019surfacestress} used to calculate the surface stresses in cylindrical nanoparticles undergoing finite deformation. Additionally, we also include the possibility of plastic deformation during lithiation/delithiation in our model, as considered in similar studies done previously \cite{anand2012cahn,cui2012finite,chakraborty2015influence,chakraborty2015combining,dora2019stress,srijan2018JAP}. The rest of the paper is organized as follows. In Section \ref{model} we set up the model geometry to be used in the analysis. Section \ref{math} has two sections: first, Section \ref{chem-elec} which covers the chemical and electrochemical aspects, and defines the modelling conditions for galvanostatic and potentiostatic charging conditions; second, Section \ref{mech} which covers the mechanical aspects with special focus on the surface stress at the outer periphery, in an abridged form. In Section \ref{rnd}, we discuss the results separately for galvanostatic case (Section \ref{Grnd}) and for potentiostatic case (Section \ref{Prnd}). The qualitative resemblance of the results with those obtained by previous researchers \cite{cheng2008influence,cui2012finite,hao2012diffusion,deshpande2010modeling} validates our computational model for further study. The crucial part of this work lies in the extensive analysis of how charging condition is primarily responsible for the overall performance of the battery under decreasing particle size. Finally in Section \ref{conclusion} we summarize the important results. \\


\section{Problem Formulation} \label{sec:math}

\subsection{Model Set up} \label{model}
\begin{figure}[h!]
	\centering
	\includegraphics[width=1\textwidth]{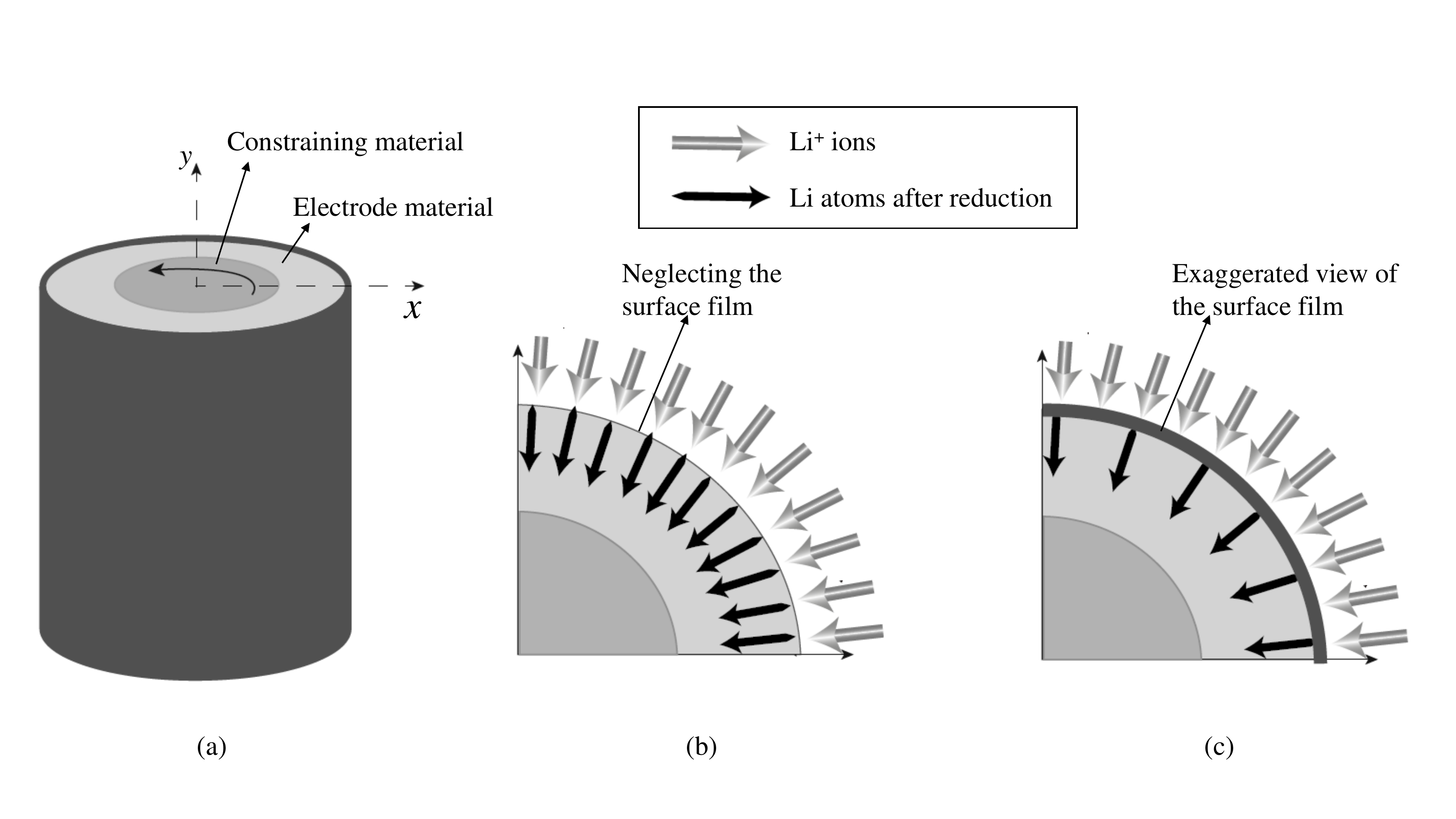}
	\caption{A schematic representation of the model set up; (b) and (c) shows the top views of a quadrant, where (b) is without and (c) is with surface-stress effects. These can be understood extensively through the results and discussions.}\label{schematic}
\end{figure}

We consider a single, cylindrical, hollow silicon anode particle in the cylindrical ($r, \theta, z$) coordinate system, having inner radius $R_{\rm i}$, outer radius $R_0$, and height $L_0$, undergoing axisymmetric deformation due to axisymmetric lithiation and delithiation. The inner core has a constraining material of radius $R_{\rm i}$, whose material properties are defined by the thickness of the core ($r_d=R_{\rm i}/R_0$) and its yield strength $\sigma_{f2}$; rest of the properties are similar to that of the electrode. The Li influx occurs radially through the outer periphery. Deformation in the axial direction is allowed. We consider a coupled-phenomenon of diffusion-induced stress and stress-enhanced diffusion. The particle is free to grow in the axial direction, keeping the radial symmetry intact. A schematic diagram depicting the model is shown in Fig. \ref{schematic}. 

\subsection{Mathematical Modelling} \label{math}
The governing differential equations characterizing the chemo-mechanical phenomenon of lithiation and de-lithiation of silicon anode particles are stated in the following sub-sections, covering the chemical and electrochemical aspects first, and then the mechanical aspect. The evolution of diffusion-induced stresses have been studied extensively by previous researchers  \cite{chakraborty2015combining,bower2011finite,cui2012finite,di2015diffusion, srijan2018JAP, dora2019stress}. Because of the availability of detailed literature, we introduce the important expressions and equations in their non-dimensionalized form, where the scheme of non-dimensionalization is presented below. 

\begin{equation} \label{nondim}
\centering
\begin{split}
\tr=\frac{r}{R_0}, \quad \tz=\frac{z}{L_0}, \quad \tD=\frac{D}{D_0}, \quad \ttime=\frac{D_0}{R_0^2}t, \quad \tu=\frac{u}{R_0}, \quad \tw=\frac{w}{L_0}, \tJ_{r,0}=\frac{V_m^{\rm Si}R_0}{D_0}J_{r,0},\\ \quad \tsigma_{r,\theta,z,{\rm eff}}=\frac{V_m^{\rm Si}}{R_{\rm g}T}\sigma_{r,\theta,z,{\rm eff}}, \quad \tsigma^0_{r,\theta,z,{\rm eff}}=\frac{V_m^{\rm Si}}{R_{\rm g}T}\sigma_{r,\theta,z,{\rm eff}}, \quad \tmu_{0,s}=\frac{1}{R_{\rm g}T}\mu_{0,s}.
\end{split}
\end{equation}
\\Here $r$ and $z$ are the radial and axial coordinates of any point in the reference configuration; ($u,v,w$) are the displacements in radial, circumferential, and axial directions, respectively (note that axisymmetric diffusion and deformation leads to axisymmetric growth, and hence, $v\equiv0$); $J_r$ and $J_0$ are the flux of lithium at any radial position $r$ and at the surface $r=R_0$; $\sigma_{r,\theta,z}$ are the Cauchy stresses in radial, cicumferential, and axial directions, respectively; $\sigma^0_{r,\theta,z}$ are the Piola-Kirchhoff stresses (PK1) in radial, cicumferential, and axial directions, respectively; lastly, $\mu$ is the chemical potential. The constants $D_0, V_m^{\rm Si}, R_{\rm g}, T$ are part of the list of mechanical properties and parameters given in Table \ref{tab:table1}. 

\begin{table}
	\caption{Mechanical properties and parameters used in the model}
	\label{tab:table1}       
	\begin{tabular}{l l}
		\hline\noalign{\smallskip}
		Material property or parameter & Value  \\
		\noalign{\smallskip}\hline\noalign{\smallskip}
		$A_0$, parameter used in activity constant & $-29549$ Jmol$^{-1}$  \cite{cui2012finite} \\
		$B_0$, parameter used in activity constant & $-38618$ Jmol$^{-1}$  \cite{cui2012finite} \\
		$D_0$, diffusivity of Si & $1 \times 10^{-16}$m$^2$s$^{-1}$  \cite{liu2011anisotropic} \\
		$\dot{d_0}$, characteristic strain rate for plastic flow in Si & $ 1 \times 10^ {-3}$ s$^{-1}$  \cite{cui2012finite} \\
		$Y_0$, modulus of elasticity of pure Si & $90.13$ GPa { \cite{rhodes2010understanding}} \\
		$m$, stress exponent for plastic flow in Si & $4$  \cite{bower2011finite} \\
		$R_{\rm g}$, universal gas constant & $8.314$ JK $^{-1}$ mol$^{-1}$ \\
		$R_0$, initial radius of unlithiated Si electrode & $50$ nm, $200$ nm \\
		$T$, temperature & $300$ K \\
		$V_m^{\rm Si}$, molar volume of Si & $1.2052 \times 10^{-5}$ m$^3$ mol$^{-1}$  \cite{cui2012finite} \\
		$x_{\rm{max}}$, maximum concentration of of Li in Si & 4.4 \\
		$\alpha$, coefficient of diffusivity & $0.18$  \cite{haftbaradaran2011continuum} \\
		$\eta$, coefficient of compositional expansion & $0.2356$  \cite{cui2012finite} \\
		$\eta_E$, rate of change of modulus of elasticity\\ with concentration & $-0.1464$  \cite{rhodes2010understanding}\\
		$\nu$, Poisson's ratio of Si & $0.28$  \cite{cui2012finite} \\
		$\sigma_f$, initial yield stress of Si & $0.49 \pm 0.08$ GPa  \cite{bucci2014measurement}\\
		$\lambda_{\rm s}$, surface Lam\'e constant & 3.5 Nm$^{-1}$  \cite{sharma2003effect} \\
		$\mu_{\rm s}$, surface Lam\'e constant & -6.23 Nm$^{-1}$  \cite{sharma2003effect} \\
		$k_0$, reaction rate constant & $1 \times 10^{-12}$ m$^{5/2}$mol$^{-1/2}$s$^{-1}$  \cite{lu2018reaction}\\
		$F$, Faraday's constant & $96500$ Cmol$^{-1}$\\
		$\phi_{\rm ref}$, reference potential & 0.88 V  \cite{di2015diffusion}\\
		$x_{\rm Li^+}$, Li ion concentration in electrolyte & 1000 molm$^{-3}$  \cite{lu2018reaction}\\
		\noalign{\smallskip}\hline
	\end{tabular}
\end{table}

\subsubsection{Chemical and Electrochemical aspect} \label{chem-elec}
The lithiation/delithiation into/ out of the anode particle is governed by the conservation equation:
\begin{equation} \label{diffeq}
\frac{\partial c}{\partial \ttime}=-\frac{\partial \tJ_r}{\partial \tr} - \frac{\tJ_r}{\tr}.
\end{equation}
Here the flux of Li ($\tJ_r$) is given as 
\begin{equation} \label{Jr}
\tJ_r= -\tilde{D}c \frac{\partial \tmu}{\partial \tr},
\end{equation}
where $\tilde{D}={\rm exp}\left(\frac{\alpha V_m^{\rm Si} \sigma^0_\theta}{R_{\rm g} T}\right)$ is the non-dimensionalized diffusivity and $c$ is the concentration of Li. The chemical potential $\tmu$ is decomposed into stress-independent and stress-dependent parts as:
\begin{equation} \label{mufunction}
\tmu=\underbrace{\frac{\mu_0^0}{R_{\rm{g}}T}+\rm{ln}(\gamma c)}_{\text{stress-independent}}+\underbrace{\tmu_{S1}+\tmu_{S2}+\tmu_{S3}}_{\text {stress-dependent}},
\end{equation}
where 
\begin{equation} \label{gamma}
\gamma=\frac{1}{(1-c)},
\end{equation}
and the expressions of the terms $\tmu_{S1}, \tmu_{S2}, \tmu_{S3}$ have been properly derived in Eq. (12)-(25) of  \cite{cui2012finite}. The expression for $\gamma$ as described in Eq. \eqref{gamma} is chosen for a reason, as explained in the Appendix.

\paragraph{Initial and boundary conditions involving the constraining material} 
Equation \eqref{diffeq} is solved both for the constraining material as well as the electrode material (here, silicon). In both these domains, it needs an initial condition and two boundary conditions. We start with a Li-free particle, thus $c(\tr,0)=0$ in both the domains. At the centre of the cylinder, the flux is always zero. Therefore, $\tJ_r(0,\ttime)=0$ at the centre. At the interface of the constraining material and the electrode, we have continuity of concentration and flux. Due to very low coefficient of volumetric expansion of the constraining material, the concentration of Li within it and its expansion in volume are low. The concentration levels also depend upon the diffusivity of the material, although in the present model, diffusivities in both the domains are considered to be the same. The boundary condition at the outer periphery is determined by the method of charging the battery: first, galvanostatic charging condition where the C-rate is kept constant; second, potentiostatic charging condition where the interface voltage drop remains constant. We elaborate on these charging conditions in the following sections. 

\paragraph{Interface voltage-drop} \label{voltdrop}
We know, while charging, the Li ions in the electrolyte reduces to form Li atoms at the surface of the anode, before diffusion. Thus, the electrochemical reaction at the surface of the anode can be written as  \cite{di2015diffusion} 
\begin{equation} \label{echemreaction}
{\rm e^{-}}\rvert_{\text{electrode surface}}\;+\;{\rm Li^+}\rvert_{\text{electrolyte}} \xtofrom[\text{discharging}]{\text{charging}} {\rm Li}\rvert_{\text{electrode surface}}
\end{equation}

Let $\bar{\mu}_{\rm Li}$, $\bar{\mu}_{\rm Li^+}$, and $\bar{\mu}_{\rm e^-}$ denote the electrochemical potentials of Li atoms (on electrode surface), Li$^+$ ions (in the electrolyte), and electrons (on electrode surface). We define the \textit{driving force}, commonly known as the `overpotential' for the chemical reaction (Eq. \eqref{echemreaction}), by taking the difference between the electrochemical potentials of the reactants and products, divided by Faraday's constant, given as
\begin{equation} \label{charging}
\eta_o=\frac{1}{F}\left[\bar{\mu}_{\rm Li}-(\bar{\mu}_{\rm Li^+}+\bar{\mu}_{\rm e^-}) \right],
\end{equation}
for charging, and
\begin{equation} \label{discharging}
\eta_o=\frac{1}{F}\left[(\bar{\mu}_{\rm Li^+}+\bar{\mu}_{\rm e^-})-\bar{\mu}_{\rm Li} \right],
\end{equation}
for discharging. The nature of the overpotential drives the reaction towards a particular direction, as follows:\\
- when $\eta_o>0$, Li oxidizes to form Li$^+$,\\
- when $\eta_o<0$, Li$^+$ reduces to form Li, and\\
- when $\eta_o=0$, the reaction is in equilibrium condition.\\

Further, we may write the electrochemical potentials  $\bar{\mu}_{\rm Li}$, $\bar{\mu}_{\rm Li^+}$, and $\bar{\mu}_{\rm e^-}$, in terms of chemical potentials and electric potentials as
\begin{equation} \label{echempot}
\bar{\mu}_i=\mu_i+z_iF\phi_i,
\end{equation}
where $\mu_i$ is the chemical potential of the $i$th charged species, $z_i$ is its valence, and $\phi_i$ is the electric potential field where the charged species $i$ exists. In our case, the electrochemical potential of each species is expressed as 
\begin{subequations} \label{echempot1}
 \begin{align}
\bar{\mu}_{\rm Li}&=\mu_{\rm Li},\\
\bar{\mu}_{\rm Li^+}&=\mu_{\rm Li^+}+F\phi_{\rm e},\\
\bar{\mu}_{\rm e^-}&=\mu_{\rm e^-}-F\phi_{\rm a}.
 \end{align}
\end{subequations}
Here, Li atoms are neutral species, Li$^+$ ions have a valence of (+1), and e$^-$ have a valence of ($-$1); $\phi_{\rm e}$ and $\phi_{\rm a}$ are the electric potentials in the electrolyte and anode, respectively.\\

At this point, we make two important assumptions:\\

First, the electrolyte and the electrode are assumed to be infinite reservoirs of Li$^+$ ions and electrons, respectively, and their activities are equal to 1. We know, the chemical potential of any species is given by $\mu_i=\mu_i^0+R_{\rm g}T{\rm ln}(a_i)$, where $a_i$ is the \textit{activity} of the species, and $\mu_i^0$ is the reference chemical potential. Therefore in Eq. \eqref{echempot1}, $\mu_{\rm Li^+}=\mu_{\rm Li^+}^0$ and $\mu_{\rm e^-}=\mu_{\rm e^-}^0$, but $\mu_{\rm Li}=\mu_{\rm Li}^0+\hat{\mu}_{\rm Li}$, where $\hat{\mu}_{\rm Li}$ is a function of concentration, stress, and temperature (refer Eq. \eqref{mufunction}).\\

Second, the electric potentials $\phi_{\rm e}$ and $\phi_{\rm a}$ are assumed to be constant within the electrolyte and the anode. The variations occur only in the electrode-electrolyte surface. \\
\\Based on the above assumptions, we re-write Eq. \eqref{echempot1} as

\begin{subequations} \label{eq3.5}
\begin{align}
\bar{\mu}_{\rm Li}&=\mu_{\rm Li}^0+\hat{\mu}_{\rm Li},\\
\bar{\mu}_{\rm Li^+}&=\mu_{\rm Li^+}^0+F\phi_{\rm e},\\
\bar{\mu}_{\rm e^-}&=\mu_{\rm e^-}^0-F\phi_{\rm a}.
\end{align}
\end{subequations}

Now using Eq. \eqref{eq3.5} in \eqref{charging} for charging, we have 
\begin{equation}\label{eq3.6}
\eta_o=\frac{\hat{\mu}_{\rm Li}}{F}-\frac{\mu_{\rm Li^+}^0+\mu_{\rm e^-}^0-\mu_{\rm Li}^0}{F}+\Delta\phi,
\end{equation}
where $\Delta\phi=\phi_{\rm a}-\phi_{\rm e}$ is the ``voltage drop" across the electrode-electrolyte interface.\\
At equilibrium, $\eta_o=0$, thus from Eq. \eqref{eq3.6} we have the equilibrium interfacial voltage drop
\begin{equation} \label{eq3.7}
\Delta\phi_{\rm eq}=\phi_{\rm ref}-\frac{\hat{\mu}_{\rm Li}}{F},
\end{equation}
where $\displaystyle \phi_{\rm ref}=\frac{\mu_{\rm Li^+}^0+\mu_{\rm e^-}^0-\mu_{\rm Li}^0}{F}$ is the reference electric potential.\\
From Eq. \eqref{eq3.6} and \eqref{eq3.7} we can write
\begin{equation} \label{eq3.8}
\eta_o=\Delta\phi-\Delta\phi_{\rm eq}.
\end{equation}

The overpotential $\eta_o$ is related to the net current density $I_{\rm n}$ (refer Eq. \eqref{eq3a}) in a non-equilibrium electrochemical reaction by the well-known Butler-Volmer equation,
\begin{equation} \label{eq3.9}
I_{\rm n}=I_0\left[{\rm exp}\left(-\beta\frac{F\eta_o}{R_{\rm g}T}\right)-{\rm exp}\left((1-\beta)\frac{F\eta_o}{R_{\rm g}T}\right)\right],
\end{equation}
where $I_0$ is the concentration-dependent exchange current density, given as 
\begin{equation} \label{eq3.10}
I_0=Fk_0x_{\rm m}x_{\rm Li^+}^{(1-\beta)}(1-c_s)^{\beta}c_s^{(1-\beta)}.
\end{equation}
Here $c_s$ is the Li concentration at the anode surface, $x_{\rm Li^+}$ is the Li-ion concentration in the electrolyte, $k_0$ is the heterogeneous reaction rate constant, and $\beta$ is taken to be 0.5.

\subparagraph{Galvanostatic condition} \label{Gcharge}
In case of galvanostatic charging/discharging of the battery, we decide a constant C-rate $N$, and the anode particles are charged/discharged with a constant current density $I_{\rm n}$, given as \cite{lu2018reaction} 
\begin{equation} \label{eq3a}
I_{\rm n} = \frac{FR_0x_{\rm m}}{2}\frac{N}{3600},
\end{equation}
where $F$ is Faraday's constant, and $x_{\rm m}=x_{\rm max}/V_m^{\rm Si}$ is the maximum moles of Li per molar volume of Si. The flux density at the cylindrical surface $\tJ_0$ is related to $I_{\rm n}$ as \cite{lu2018reaction}
\begin{equation} \label{eq3b}
\tJ_0=\frac{I_{\rm n}R_0}{FD_0x_{\rm m}}.
\end{equation}
It is important to note that it is this relation which is used as the second boundary condition for the governing differential equation for diffusion (Eq. \eqref{diffeq}).\\

Using Eq. \eqref{eq3.9} we find out the overpotential as
\begin{equation} \label{eq3.11}
\eta_o=2\frac{R_{\rm g}T}{F}{\rm sinh^{-1}}\left(-\:\frac{I_{\rm n}}{2I_0}\right)
\end{equation}
for charging, and
\begin{equation} \label{eq3.12}
\eta_o=2\frac{R_{\rm g}T}{F}{\rm sinh^{-1}}\left(\frac{I_{\rm n}}{2I_0}\right)
\end{equation}
for discharging.\\
Therefore from Eq. \eqref{eq3.8}, we have the interfacial voltage drop as 
\begin{equation} \label{eq3.13}
\Delta\phi=\phi_{\rm ref}-\frac{\hat{\mu}_{\rm Li}}{F}+2\frac{R_{\rm g}T}{F}{\rm sinh^{-1}}\left(\pm \:\frac{I_{\rm n}}{2I_0}\right)
\end{equation}
The interfacial voltage drop, being dependent upon various factors such as stress, concentration, particle-size, temperature, C-rate, etc. is sensitively modulated when any of these factors vary. In the Results and Discussion section we analyze the effects of particle size and C-rate on $\Delta\phi$.

\subparagraph{Potentiostatic condition} \label{Pcharge}
In case of potentiostatic charging/discharging of the battery, the charging/discharging occurs at a constant interface potential difference $ \Delta \phi$ (Eq. \eqref{eq3.6}), whereas the charging/discharging rate of Li may vary. In the present case, we set the interface potential difference $\Delta \phi$ at 5.31 mV \cite{lu2018reaction}. \\ \\
From the Butler-Volmer equation (Eq. \eqref{eq3.9}) the constant current density $I_{\rm n}$ at which the anode particles are charged/discharged is given as 
\begin{equation} \label{eq3.14}
I_{\rm n}=-2I_0 {\rm sinh}\left[\frac{F}{2R_{\rm g}T}(\Delta \phi - \Delta \phi_{\rm eq})\right]
\end{equation}

\subsubsection{Mechanical aspect} \label{mech}
Pertaining to the axisymmetric situation, the radial component of the mechanical equilibrium equation in the reference configuration is given as: 
\begin{equation} \label{mecheq}
\frac{\partial \tsigma_{r}^{0}}{\partial \tr} + \dfrac{\tsigma_{r}^{0}-\tsigma_{\theta}^{0}}{\tr} = 0,
\end{equation}
Considering linear-elastic, isotropic behaviour, the strain-energy density function in the reference configuration, assuming small  deformation in the elastic domain, takes the form:
\begin{equation} \label{SEDF}
W=\frac{J^c}{2}\frac{\tilde{Y}}{(1+\nu)}\left[\frac{\nu}{1-2\nu}(E^e_{kk})^2+E^e_{jk}E^e_{kj}\right],
\end{equation}
where $\tilde{Y}=\tilde{Y}_0(1+\eta_E x_{\rm max}c)$ is the concentration-dependent modulus of elasticity; $\nu$ is the Poisson's ratio; $J^c=1+3\eta x_{\rm max}c$ gives the ratio of final to initial volumes related to the stress-free, unconstrained swelling/contraction behaviour of the Si particle on diffusion, in/out of the anode, respectively. Here $x_{\rm max}$ is the maximum number of moles of Li per mole of Si ($x_{\rm max}=4.4$, therefore, in Li$_x$Si, $0\le x\le4.4$) and $c=x/x_{\rm max}$ represents the spatio-temporally varying concentration of Li inside Si particle. The PK1 stresses ($\tsigma^0_r, \tsigma^0_\theta, \tsigma^0_z$) can be derived from $W$ \cite{chakraborty2015combining} and are expressed in terms of the elastic part of the Lagrangian strains ($E^e_r, E^e_\theta, E^e_z$) as:
\begin{subequations} \label{PK1volume}
	\begin{align}
	&\tsigma_{r}^{0}=J^{c}\dfrac{\tilde{Y}}{(1+\nu)(1-2\nu)}[(1-\nu)E_{r}^{e}+\nu(E_{\theta}^{e}+E_{z}^{e})]\frac{2E_{r}^{e}+1}{1+\partial \tu/\partial \tr},\\
	&\tsigma_{\theta}^{0}=J^{c}\dfrac{\tilde{Y}}{(1+\nu)(1-2\nu)}[(1-\nu)E_{\theta}^{e}+\nu(E_{r}^{e}+E_{z}^{e})]\frac{2E_{\theta}^{e}+1}{1+\tu/\tr},\\
	&\tsigma_{z}^{0}=J^{c}\dfrac{\tilde{Y}}{(1+\nu)(1-2\nu)}[(1-\nu)E_{z}^{e}+\nu(E_{\theta}^{e}+E_{r}^{e})]\frac{2E_{z}^{e}+1}{1+\partial \tw/\partial \tz}.
	\end{align}
\end{subequations}
Instead, the description of $E^e_r, E^e_\theta, E^e_z$ are connected to the kinematics of the particle \cite{chakraborty2015influence}.\\

\paragraph{Boundary conditions involving constraining material} There is no deformation at the center of the constraining material, i.e. $\tilde{u} \rvert_{\tilde{r}=0}=0$; at the interface ($x=R_{\rm i}$) there is continuity of radial displacements and radial stresses between the constraining material and silicon. In the absence of surface stress at the outer periphery of the Si particle, the radial stress $\tsigma_r \rvert_{\tr=1}=0$, since we assume a negligible effect of the electrolyte and SEI on the anode particles. However, when we consider a particle of outer radius $<$ 50 nm, the surface stresses can no longer be neglected, and in such cases, at the periphery, 
\begin{equation} \label{SSBC}
\tsigma^0_r \er = -\frac{\widehat{\tsigma}_{\theta}^0}{R_0} \frac{w_0 V_m^{\rm Si}}{R_{\rm g} T} \er,
\end{equation}
where $\widehat{\tsigma}_{\theta}^0$ is the PK1  surface stress in the circumferential direction, the mathematical formulation of which is presented in details in \cite{ASJC2019surfacestress}. There is no constraint in the axial direction, and hence the axial force over the entire domain (from $x=0$ to $x=R_0$) is zero. Mathematically,
\begin{equation}
	\underbrace{2\pi\int_{0}^{\frac{R_i}{R_o}}\tsigma_z^0\tr d\tr}_\text{core constraining material}+\underbrace{2\pi\int_{\frac{R_i}{R_o}}^{1}\tsigma_z^0\tr d\tr}_\text{outer electrode material}=0
\end{equation}


\section{Results and Discussions} \label{rnd}
The governing equations as obtained in Section 2 are highly coupled and therefore, are solved numerically using the PDE interface of COMSOL Multiphysics 5.3a simulation software. The results have been discussed in two separate sections: first, for the galvanostatic charging condition; second, for the potentiostatic charging condition. For future reference, we define the state-of-charge or SOC as the average concentration of lithium within the Si particle, with respect to its reference configuration, and is mathematically expressed as: SOC=$\displaystyle \frac{\left(\int_{0}^{R_0}c2\pi r\;dr\right)}{\left(\int_{0}^{R_0}2\pi r\;dr\right)}$. In the galvanostatic condition, we consider a complete charging-discharging cycle (charging upto 80-90 $\%$ SOC) to undertsand the electrochemical performance better. However, in the potentiostatic case, the anode particle attains an equilibrium SOC during lithiation, and no further charging increases the SOC beyond that. Detailed discussions are done in the respective sections. \\

\subsection{Galvanostatic charging condition} \label{Grnd}

Since the detailed analysis on the evolution of stresses can be found in earlier literature \cite{cui2012finite, dora2019stress}, we start our discussion with the plots analyzing the electrochemical performance of the electrode particle. \\

\begin{figure} [h!]
	\centering
	\begin{subfigure}[b]{0.5\textwidth}
		\includegraphics[width=\textwidth]{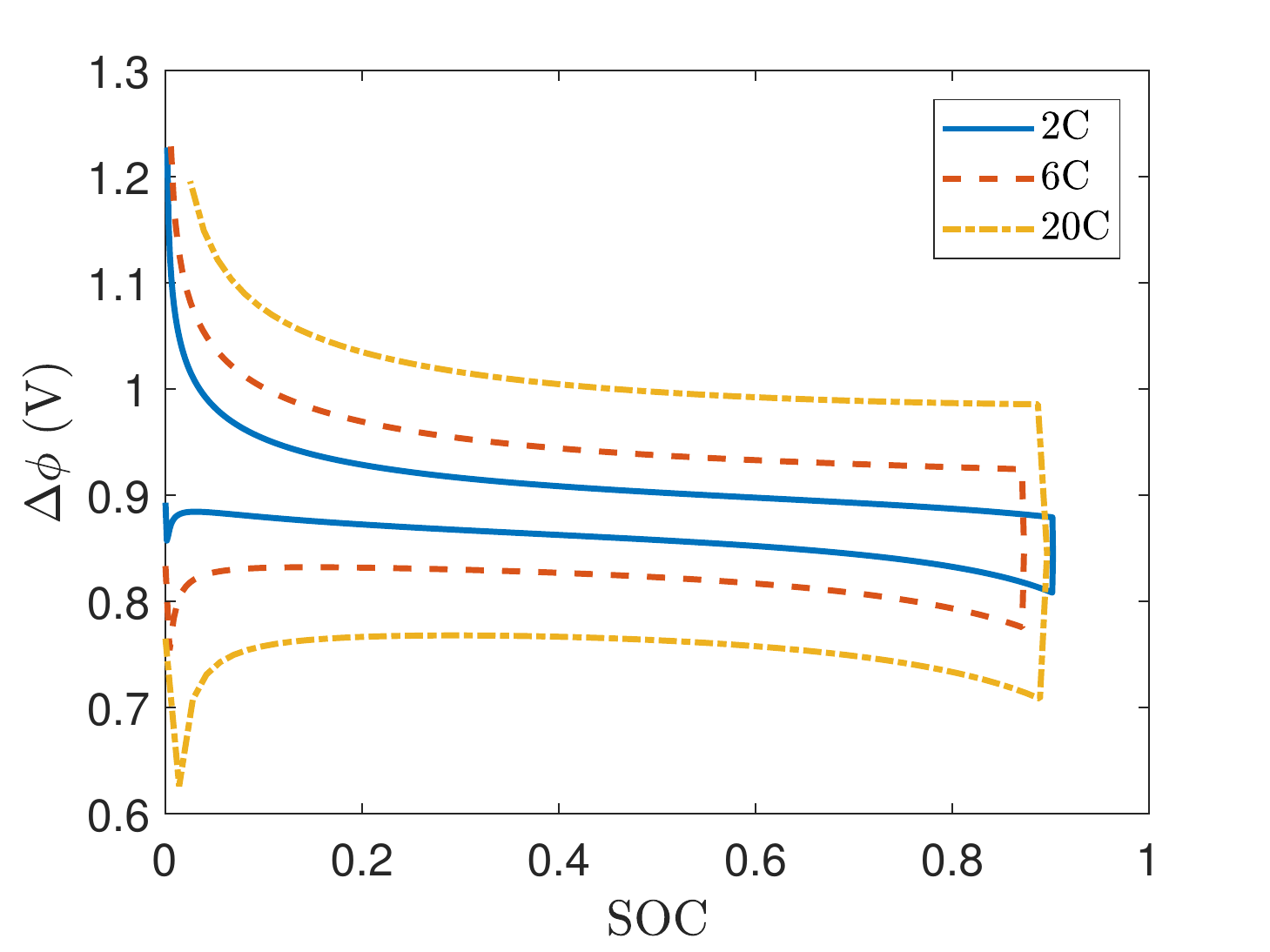}
		\caption{}
	\end{subfigure}%
	\begin{subfigure}[b]{0.5\textwidth}
		\includegraphics[width=\textwidth]{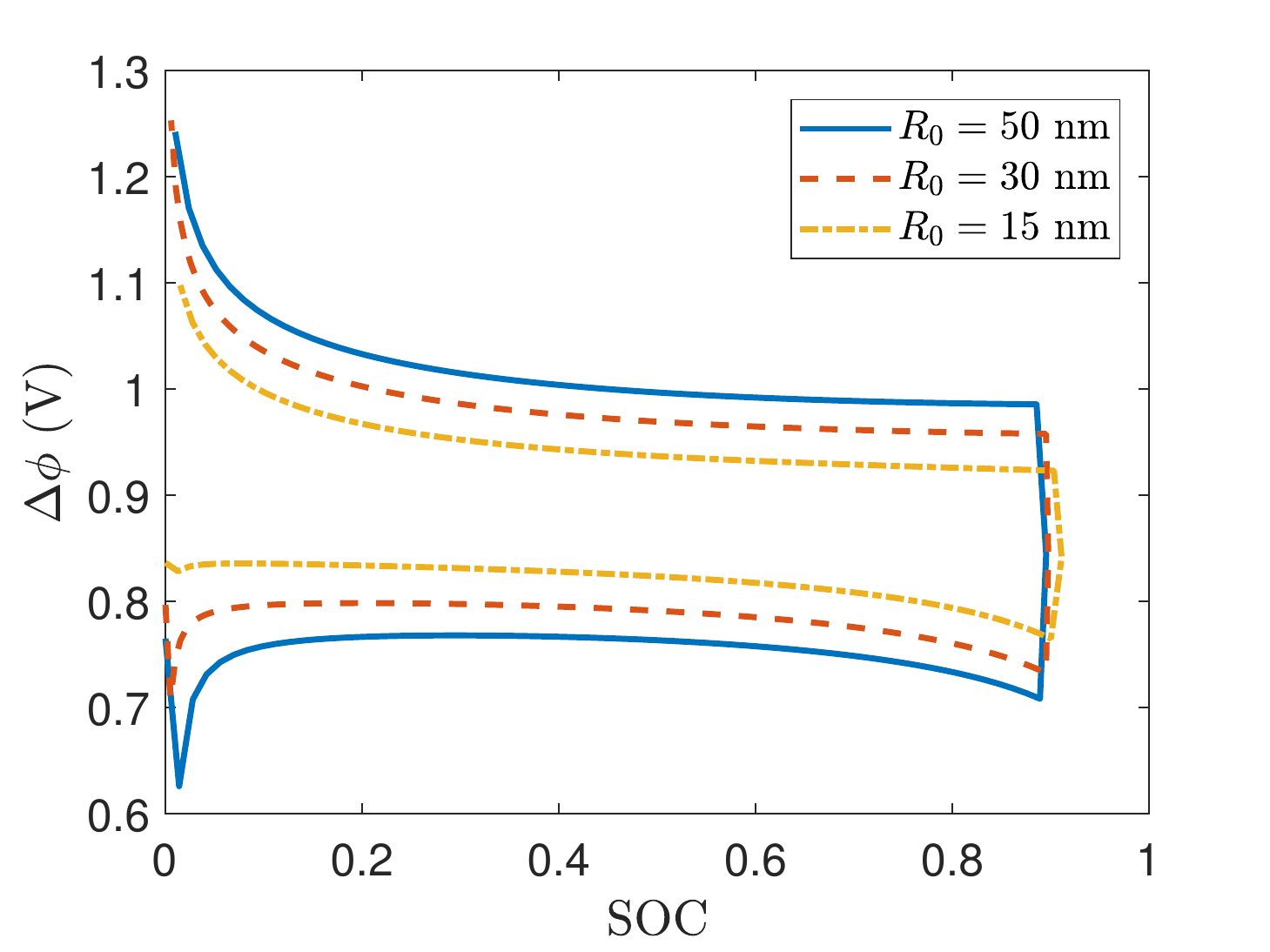}
		\caption{}
	\end{subfigure}
	\caption{Variation of interface voltage-drop with (a) C-rate (keeping particle size constant at 50 nm) and (b) particle size (keeping C-rate constant at 20C), plotted against SOC for a complete charging-discharging cycle.} \label{galvanoRCeffect}
\end{figure}

\begin{figure} [h!]
	\centering
	\begin{subfigure}[b]{0.5\textwidth}
		\includegraphics[width=\textwidth]{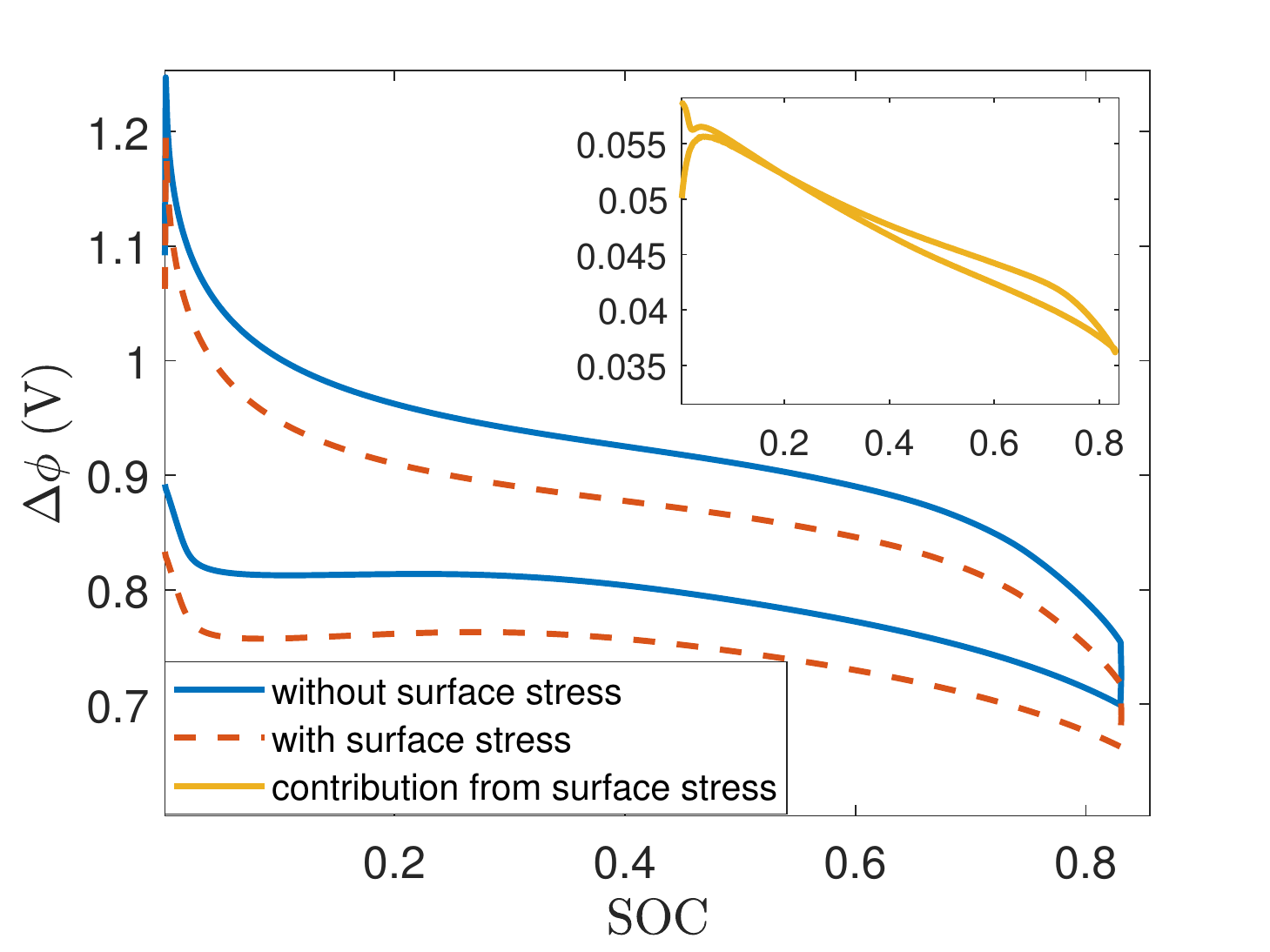}
		\caption{}
	\end{subfigure}%
	\begin{subfigure}[b]{0.5\textwidth}
		\includegraphics[width=\textwidth]{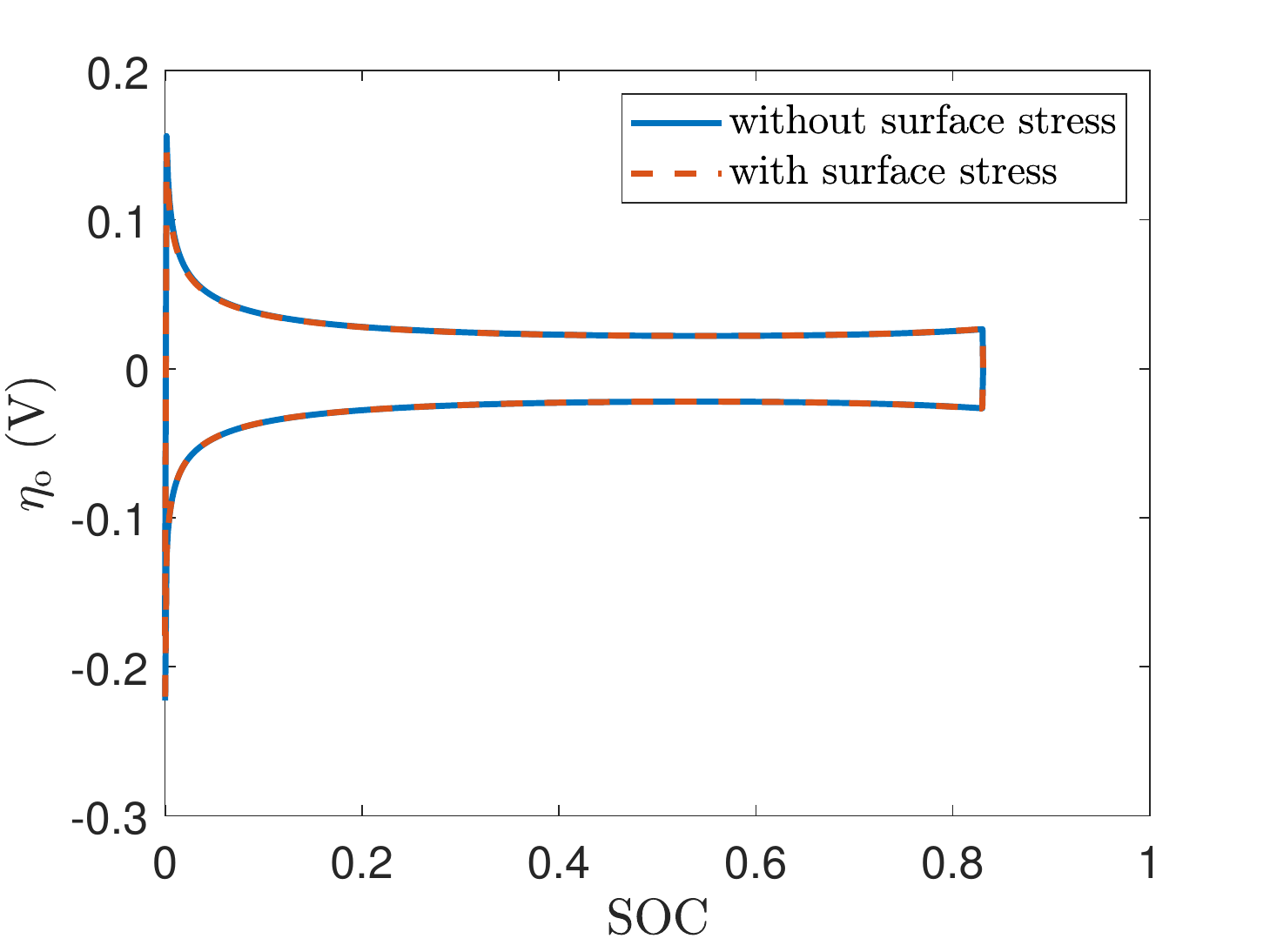}
		\caption{}
	\end{subfigure}
	\begin{subfigure}[b]{0.5\textwidth}
		\includegraphics[width=\textwidth]{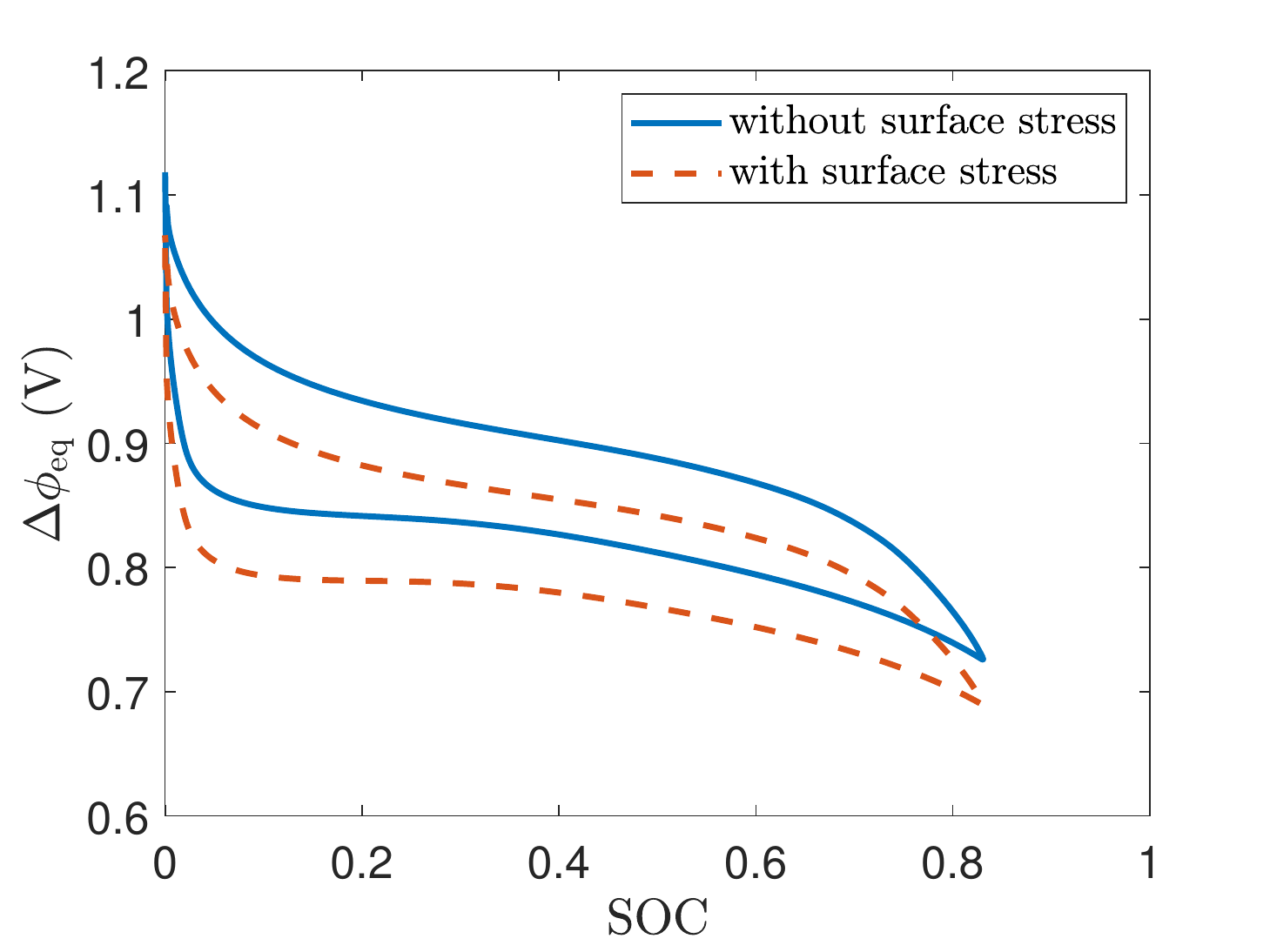}
		\caption{}
	\end{subfigure}%
	\begin{subfigure}[b]{0.5\textwidth}
		\includegraphics[width=\textwidth]{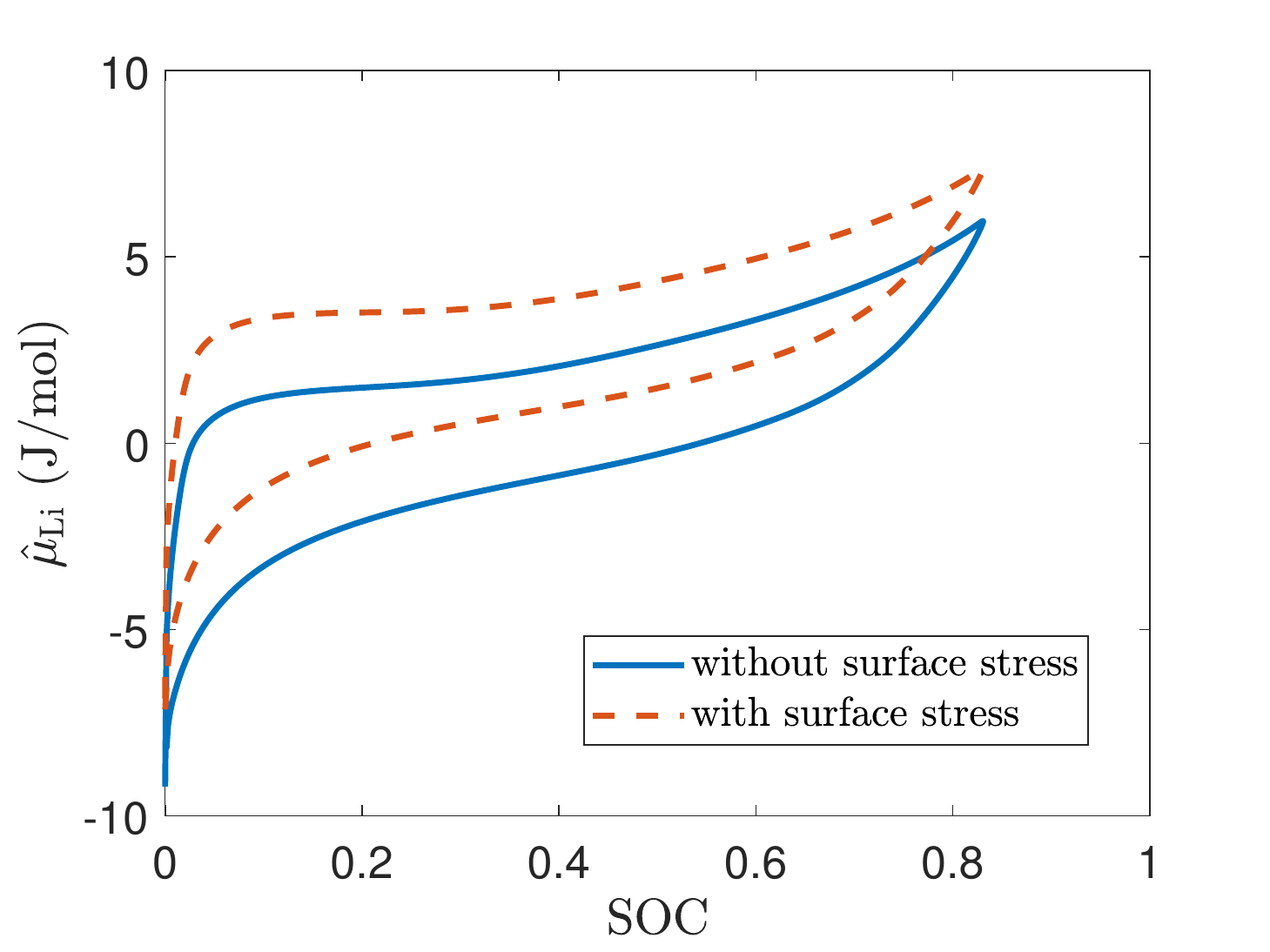}
		\caption{}
	\end{subfigure}
	\caption{Keeping C-rate constant at 20C and particle size 15 nm, we plot (a) interface voltage-drop against SOC, with and without surface stress, along with the contribution of surface stress (inset figure), (b) overpotential against SOC, with and without surface stress, (c) equilibrium potential against SOC, with and without surface stress, and (d) chemical potential at the surface of Si particle against SOC, with and without surface stress.} \label{SSeffG}
\end{figure}

The electrochemical performance in case of galvanostatic condition can be best judged by the interface voltage-drop variation. As mentioned earlier, interface voltage-drop is dependent on various factors, most important among them being particle size and charging rate. Figures \ref{galvanoRCeffect} (a) and (b) shows how a variation in particle sizes and charging rates affect the interface-voltage profiles.\\

As charging/discharging rate determines the net current density and the influx rate (Eq. \eqref{eq3a} and \eqref{eq3b}), it directly affects the overpotential value, as well as the equilibrium potential. Thus to examine the effect of C-rate on the overall voltage-drop profile, we plot $\Delta \phi$ against SOC at various C-rates (\ref{galvanoRCeffect} (a)), for a particle of radius 50 nm, considering no surface stresses. As $N$ (charging/discharging rate) increases, the net current density $I_{\rm n}$ increases, which enhances the overpotential magnitude, thus decreasing $\Delta \phi$ during charging and increasing $\Delta \phi$ during discharging. The voltage drop profile becomes broader with increasing C-rate. Next, to examine the effect of particle size on the voltage-drop profile, we plot $\Delta \phi$ against SOC for various $R_0$ values (\ref{galvanoRCeffect} (b)), keeping C-rate fixed at 20 C. It is observed that, as the particle size decreases, the net current density for a given C-rate (refer Eq. \eqref{eq3a}) decreases, which further decrease the magnitude of the overpotential. This increases the interfacial voltage drop during charging and decreases it during discharging, hence narrowing the gap in voltage-drop cycle. \\ 

\begin{figure}[h!]
	\centering
	\includegraphics[width=0.7\textwidth]{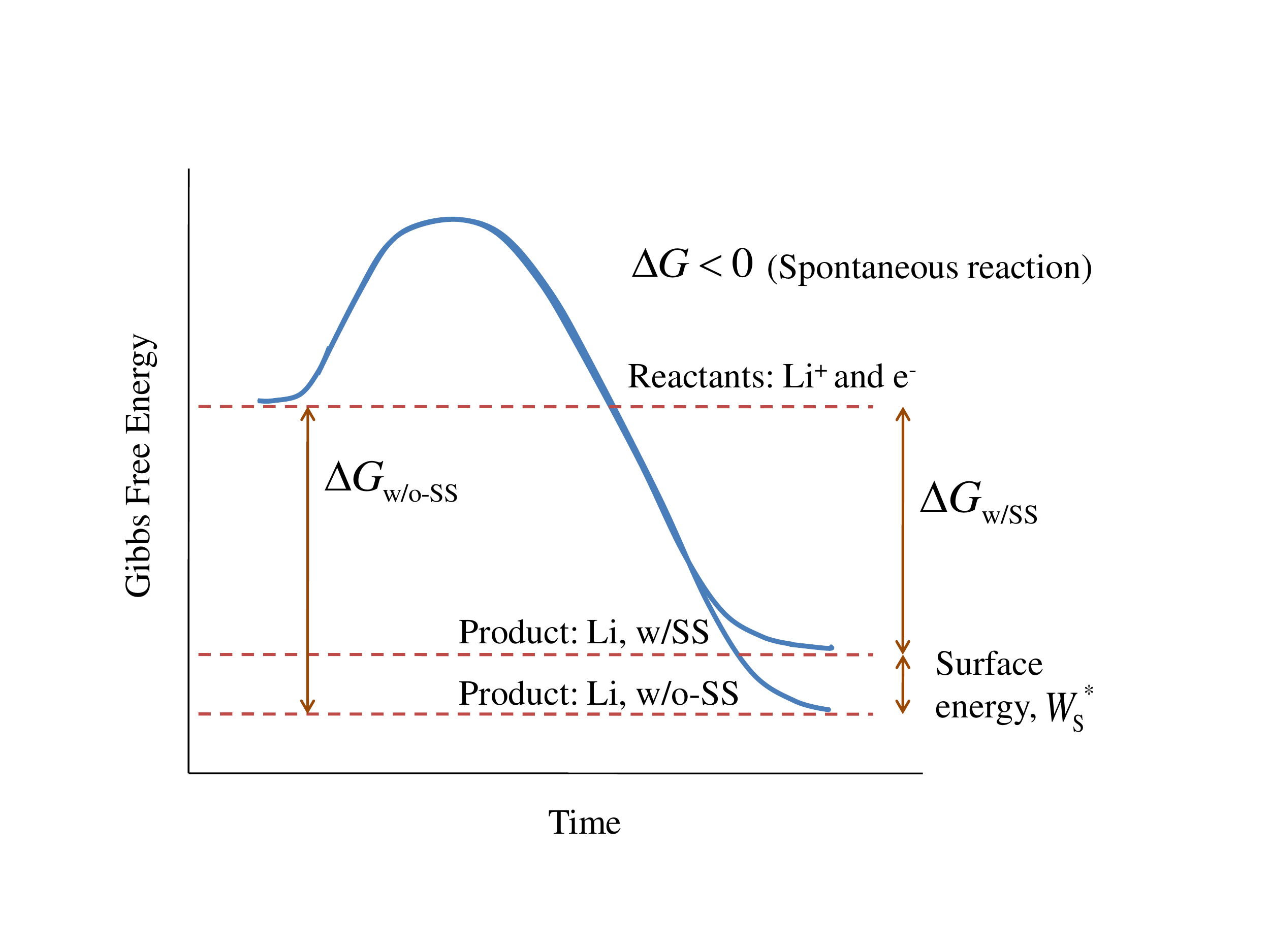}
	\caption{A schematic representation of the variation of Gibbs free energy with time, with and without considering surface stress, for the reaction: Li$^+$ + e$^-$ $\rightarrow$ Li at the surface of the anode particle. [$W^*_s$ is the interface energy per unit area in the
		reference configuration \cite{ASJC2019surfacestress}.] }\label{gibbs}
\end{figure}

The surface stress effects become more significant with decrease in the size of the anode particles \cite{cheng2008influence,deshpande2010modeling,ASJC2019surfacestress}. Therefore, to analyze the effect of surface stress on $\Delta \phi$, we plot $\Delta \phi$ against SOC (Fig. \ref{SSeffG} (a)), with and without considering surface stress, for a particle of radius 15 nm, charged/discharged at 20C rate. It is observed that the voltage drop profile shifts downward due to the effect of surface stress. In the inset, we see the contribution of surface stress in $\Delta \phi$; the shift, as we can see, is more or less constant throughout, except for a brief period when Li concentration inside anode is very low. Now, we would try to correlate this shift with the overall electrochemical performance. To understand the shift, we plot overpotential against SOC (Fig. \ref{SSeffG} (b)), with and without surface stress, equilibrium potential against SOC (Fig. \ref{SSeffG} (c)), with and without surface stress, and chemical potential at the surface of Si particle against SOC (Fig. \ref{SSeffG} (d)), with and without surface stress. Clearly, overpotential trends in both the cases of with and without surface stress, overlaps, except for very low SOCs, which means, surface stress does not affect the overpotential directly. The effect of surface stress on overpotential is indirect, through the surface concentration terms in $I_0$, and therefore, is negligible. Next, from the equilibrium potential trends, we witness a similar negative shift as observed in case of interfacial voltage-drop, which implies the shift in $\Delta \phi$ is a direct consequence of the shift in $\Delta \phi_{\rm eq}$. From Eq. \eqref{eq3.7}, we know $\displaystyle \Delta\phi_{\rm eq}=\phi_{\rm ref}-\frac{\hat{\mu_{\rm Li}}}{F}$, where the value of $\phi_{\rm ref}$ is constant. Therefore, the shift in $\Delta\phi_{\rm eq}$ occurs due to a shift in the chemical potential at Si surface, given by $\hat{\mu_{\rm Li}}$ (se Fig. \ref{SSeffG} (d)). It is seen from Fig. \ref{SSeffG} (d) that $\hat{\mu}_{\rm Li}$ becomes less negative when surface stress is considered. We know, the chemical potential is given as:
\begin{equation} \label{mu}
\mu_i=\left[\frac{\partial G}{\partial n_i}\right]_{S,T,j\ne i},
\end{equation}
where $\mu_i$ is the chemical potential of species $i$, defined as the change in Gibbs free energy ($G$) of the system due to an infinitesimal change in the number of species $i$ in the system, keeping entropy ($S$), temperature ($T$), and other species constant. We also know, that a reaction is spontaneous when $\Delta G$ is negative. The positive shift of $\hat{\mu}_{\rm Li}$ in  Fig. \ref{SSeffG} (d) implies that the change in Gibbs free energy during the reduction reaction (Li$^+$ + e$^-$ ~$\rightarrow$ ~Li) at the surface of the anode particle is less negative in the presence of surface stress, than in its absence (Fig. \ref{gibbs}). This happens because of the presence of surface energy at the anode surface, which increases the Gibbs free energy at the product side. Therefore, when surface stress comes into play, the reduction of Li$^+$ ions to form Li atoms become less spontaneous than the case without surface stress. Since the reduction reaction determines the level of charging, we can state that the charging and hence, the electrochemical performance is affected negatively in the presence of surface stress. Further, we study how this downward shift of $\Delta \phi$ can be modulated by varying the particle radius and C-rate.\\

\begin{figure} [h!]
	\centering
	\begin{subfigure}[b]{0.5\textwidth}
		\includegraphics[width=\textwidth]{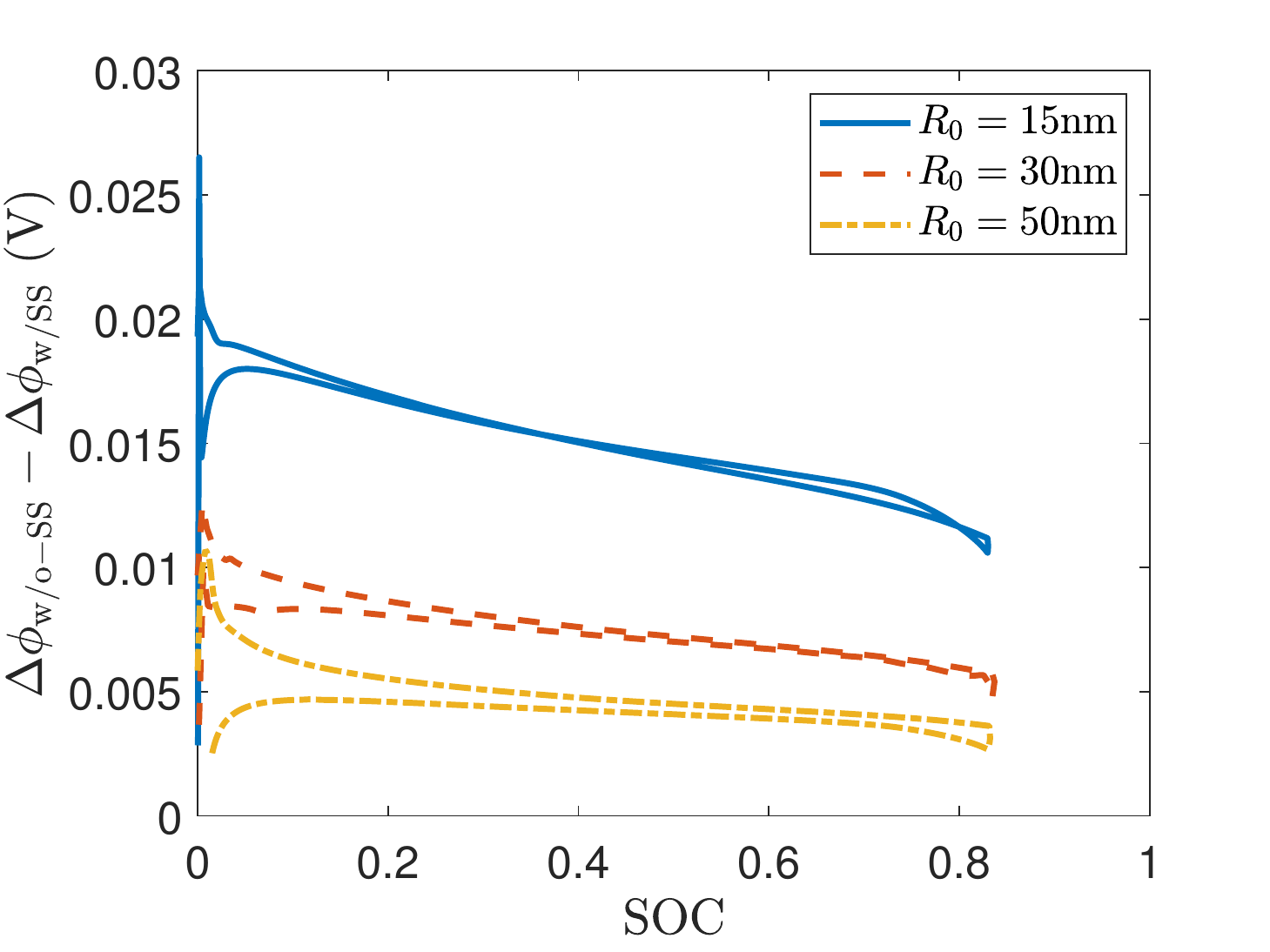}
		\caption{}
	\end{subfigure}%
	\begin{subfigure}[b]{0.5\textwidth}
	\includegraphics[width=\textwidth]{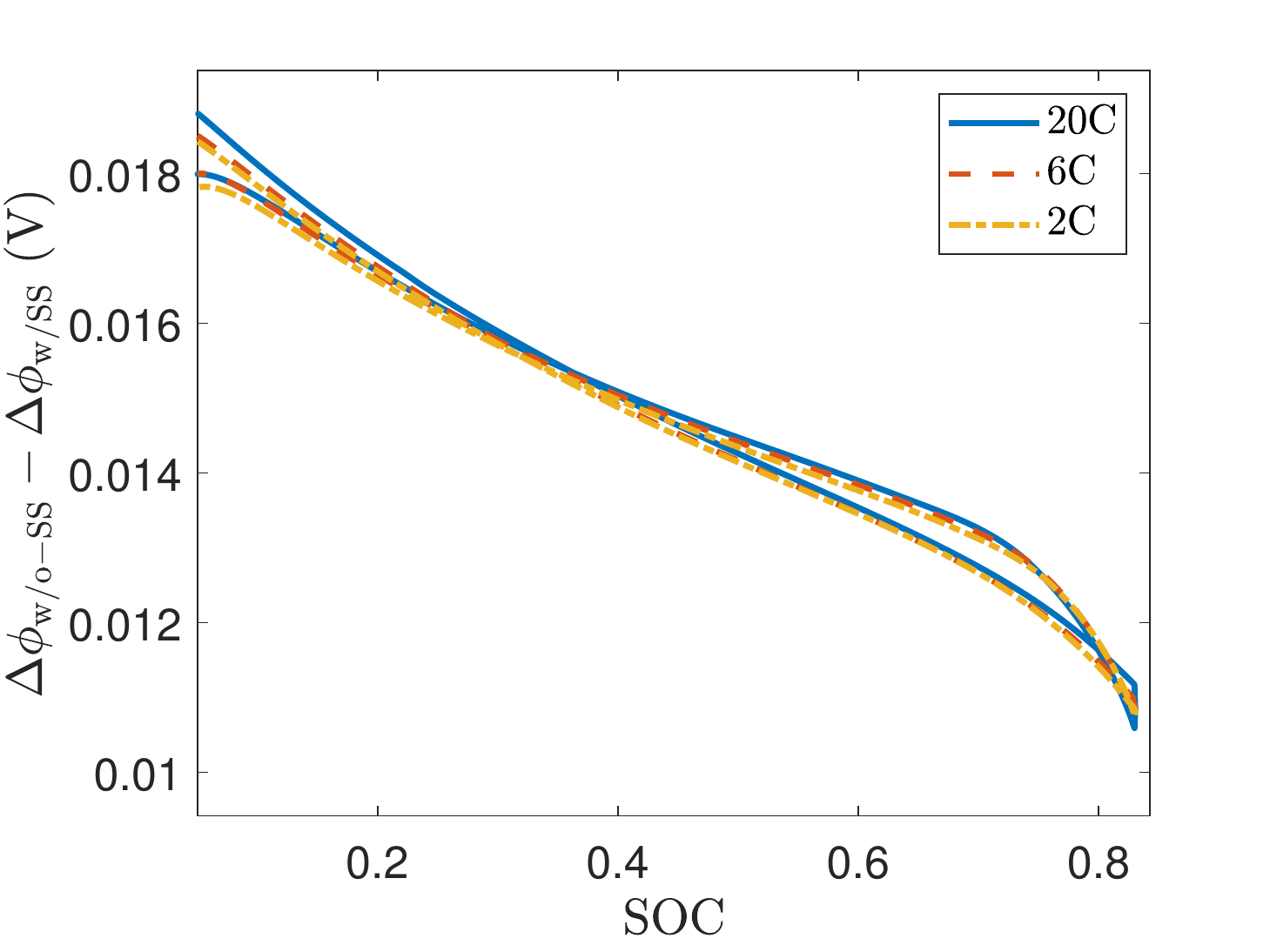}
	\caption{}
    \end{subfigure}
	\caption{Effect of surface stress varies with (a) particle size (keeping C-rate constant at 20C), and (b) C-rate (keeping particle size constant at 50 nm); plotted against SOC for a complete charging-discharging cycle.} \label{galvanoSSRCeffect}
\end{figure}

It is well-known from previous works \cite{ASJC2019surfacestress} how surface stresses and hence the DISs get modulated by the variation in particle sizes and charging rates. As mentioned earlier, interface voltage-drop is dependent on various factors, most important among them being particle size and charging rate. In Fig. \ref{galvanoSSRCeffect} (a), we observe the variation of ($\Delta \phi_{\rm w/o-SS}-\Delta \phi_{\rm w/SS}$) with SOC for different particle radii, keeping charging rate fixed at 20C for each of the cases. As the particle size decreases, the surface stress effect increases, i.e. the magnitude of negative shift as observed in Fig. \ref{SSeffG} increases. This means, as we reduce the size of the anode particle, the surface energy increases and the electrochemical performance of the particle decreases. Next, in Fig. \ref{galvanoSSRCeffect} (b), we plot the variation of ($\Delta \phi_{\rm w/o-SS}-\Delta \phi_{\rm w/SS}$) with SOC for different charging rates, keeping initial particle size same as $R_0$ = 50 nm for each of the cases. We observe that at charging rate 2C, there is no effect of surface stress, but as we increase the charging rate, the surface stress effect becomes more and more prominent. Although the overall magnitude does not vary much with increase in C-rate, the gap between charging and discharging magnitudes increases. Therefore, for very high C-rates (6C and 20C), the electrochemical performance reduces, but the reduction is comparable, and is independent of the charging rate. This result is consistent with the effects of increasing influx rate on surface stress as witnessed in an earlier study \cite{ASJC2019surfacestress}. Similar to Fig. \ref{galvanoSSRCeffect} (b), the effects of charging/influx rate on the surface stress was only visible during the initial stages of lithiation when the concentration inhomogeneity of Li is very high.\\

\begin{figure} [h!]
	\centering
	\begin{subfigure}[b]{0.5\textwidth}
		\includegraphics[width=\textwidth]{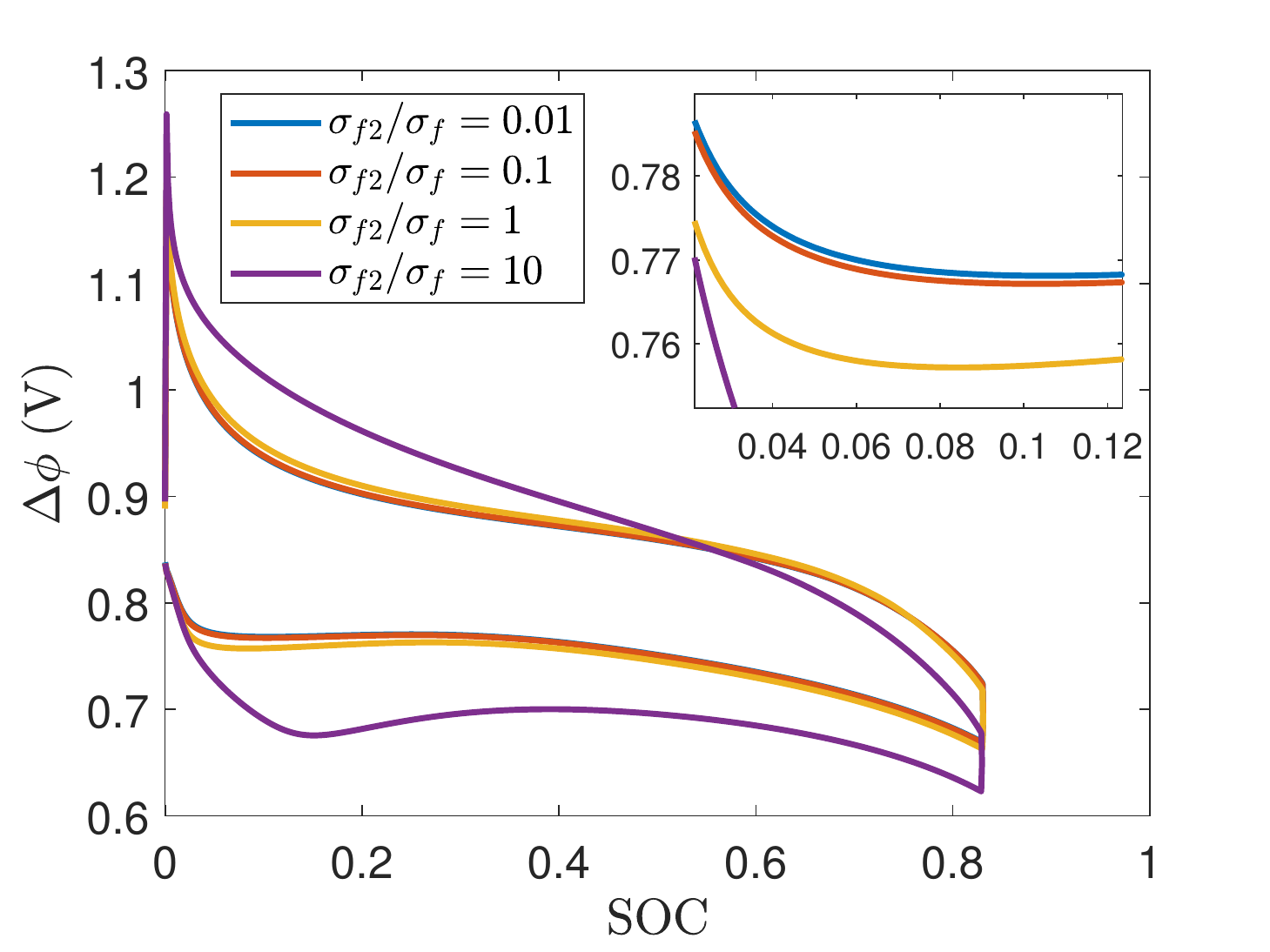}
		\caption{}
	\end{subfigure}%
	\begin{subfigure}[b]{0.5\textwidth}
		\includegraphics[width=\textwidth]{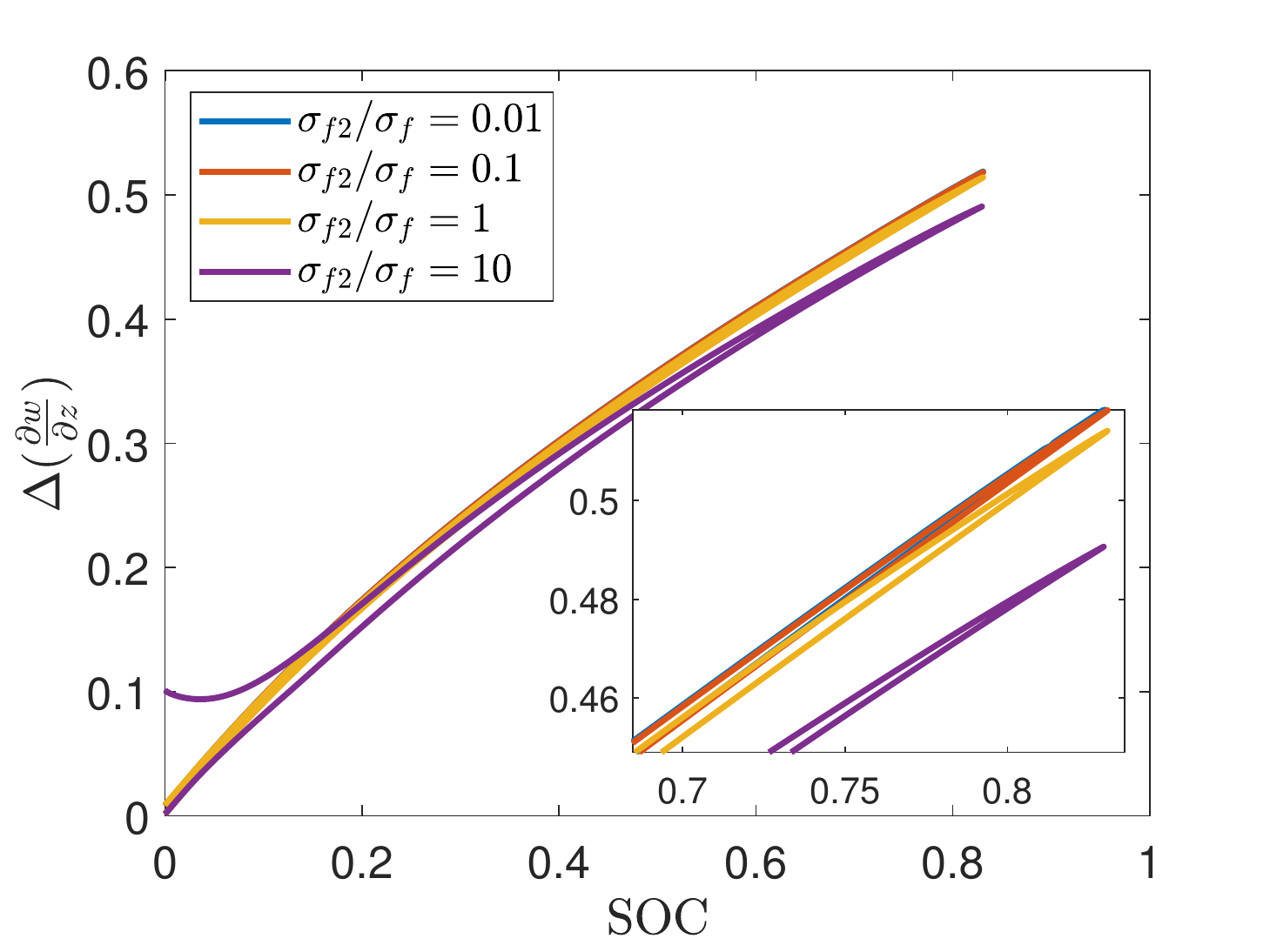}
		\caption{}
	\end{subfigure}
	\caption{(a) Variation of electrode potential, and (b) variation of length-increase over a complete charging-discharging cycle, for various yield-strengths of the constraining material. The charging rate is kept at 20C and the particle size is taken to be 15 nm.} \label{galvanosigmaeffect}
\end{figure}

In all the previous plots (Fig.\ref{SSeffG}-Fig.\ref{galvanoSSRCeffect}) the yield strengths of both the inner-core (constraining material) and the Si electrode were the same. Further, it is to be noted that the thickness of the constraining material is also kept constant at 0.3. To understand how the constraining material-properties affect the mechanical and electrochemical performances of the electrode particle, we vary the ratio of yield strengths of the constraining material to that of Si ($\frac{\sigma_{f2}}{\sigma_f}$) from 0.01 to 10 (by 4 orders of magnitude). The electrochemical performance is gauged by the electrode-potential trends, whereas the mechanical performance is reckoned by the amount of length-increase of the axially-unconstrained particle. \\

Figure \ref{galvanosigmaeffect} shows the (a) variation of electrode potential $\Delta \phi$, and (b) variation of length-increase $\frac{\partial w}{\partial z}$ over a complete charging-discharging cycle, for ratios of yield strengths $\frac{\sigma_{f2}}{\sigma_f}$ varying from 0.01 to 10. The charging (and discharging) occurs at a rate of 20C, and the particle size is taken as 15 nm in each of the cases. We have neglected the effect of surface stresses in this figure, so as to focus only on the effects of constraining material-properties. The increasing yield strength of the constraining material at the core shifts the electrode potential $\Delta \phi$ towards the downward direction (Fig.\ref{galvanosigmaeffect}(a)), indicating an increase in the surface energy and thus, decrease in the electrochemical performance of the electrode, albeit for ratios of 1 and below, this reduction is not significant. On the other hand, the length-increase of the electrode particle reduces due to an increase in the yield strength of the constraining material, from 0.01 to 10. The change in the length-increase, however, remains within 1 - 5 $\%$. It is important to note that the position of the constraining material, as well as its thickness play very crucial roles in determining the amount of length-increase. An extensive analysis on this topic exists in Figures 3 and 4 of  \cite{chakraborty2015combining}. Our motivation behind considering a constraining material at the inner-core is to understand the effects of constraining material and surface stresses, individually and simultaneously, in the same study. 

\begin{figure} [h!]
	\centering
	\begin{subfigure}[b]{0.5\textwidth}
		\includegraphics[width=\textwidth]{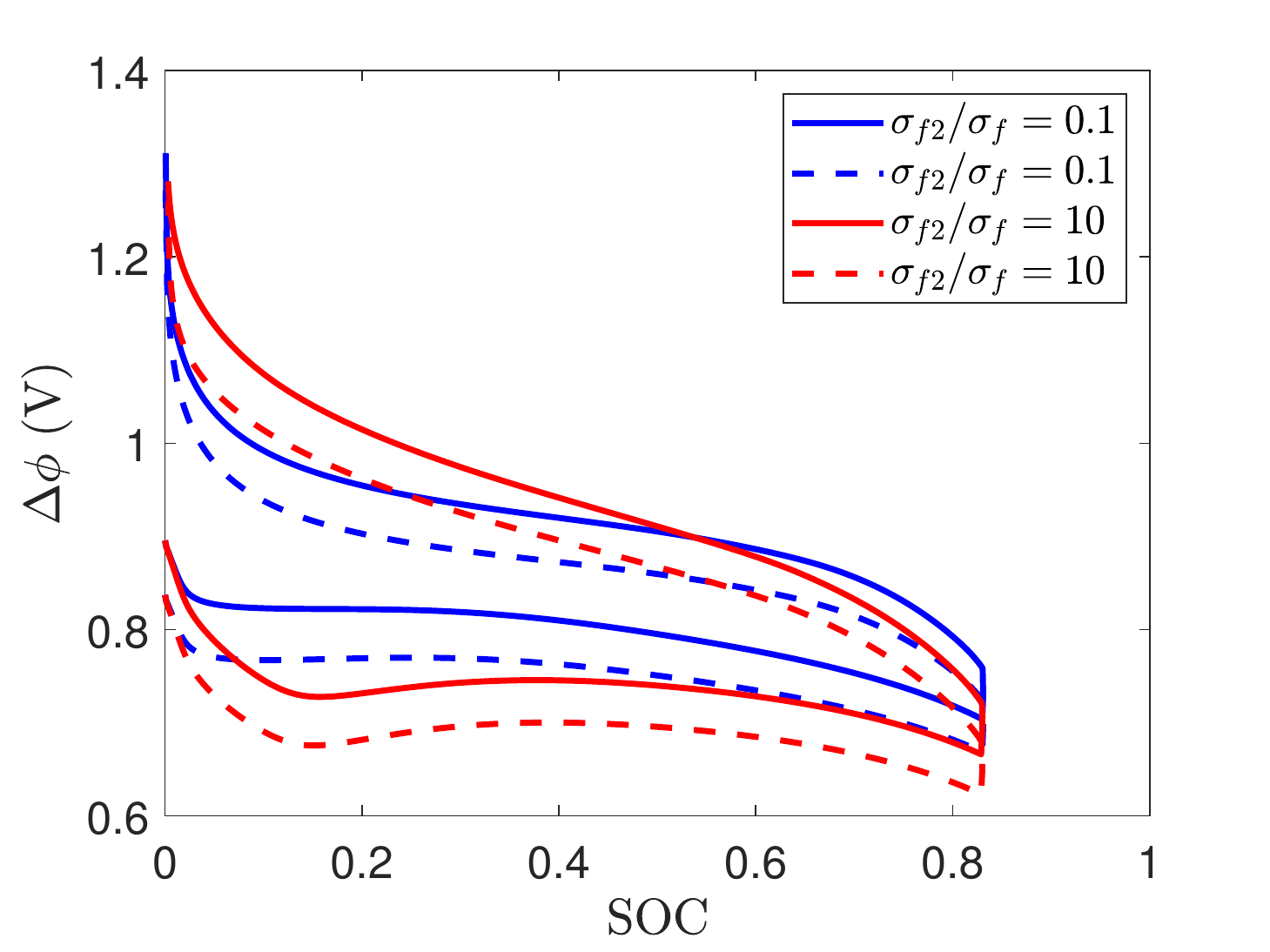}
		\caption{}
	\end{subfigure}%
	\begin{subfigure}[b]{0.5\textwidth}
		\includegraphics[width=\textwidth]{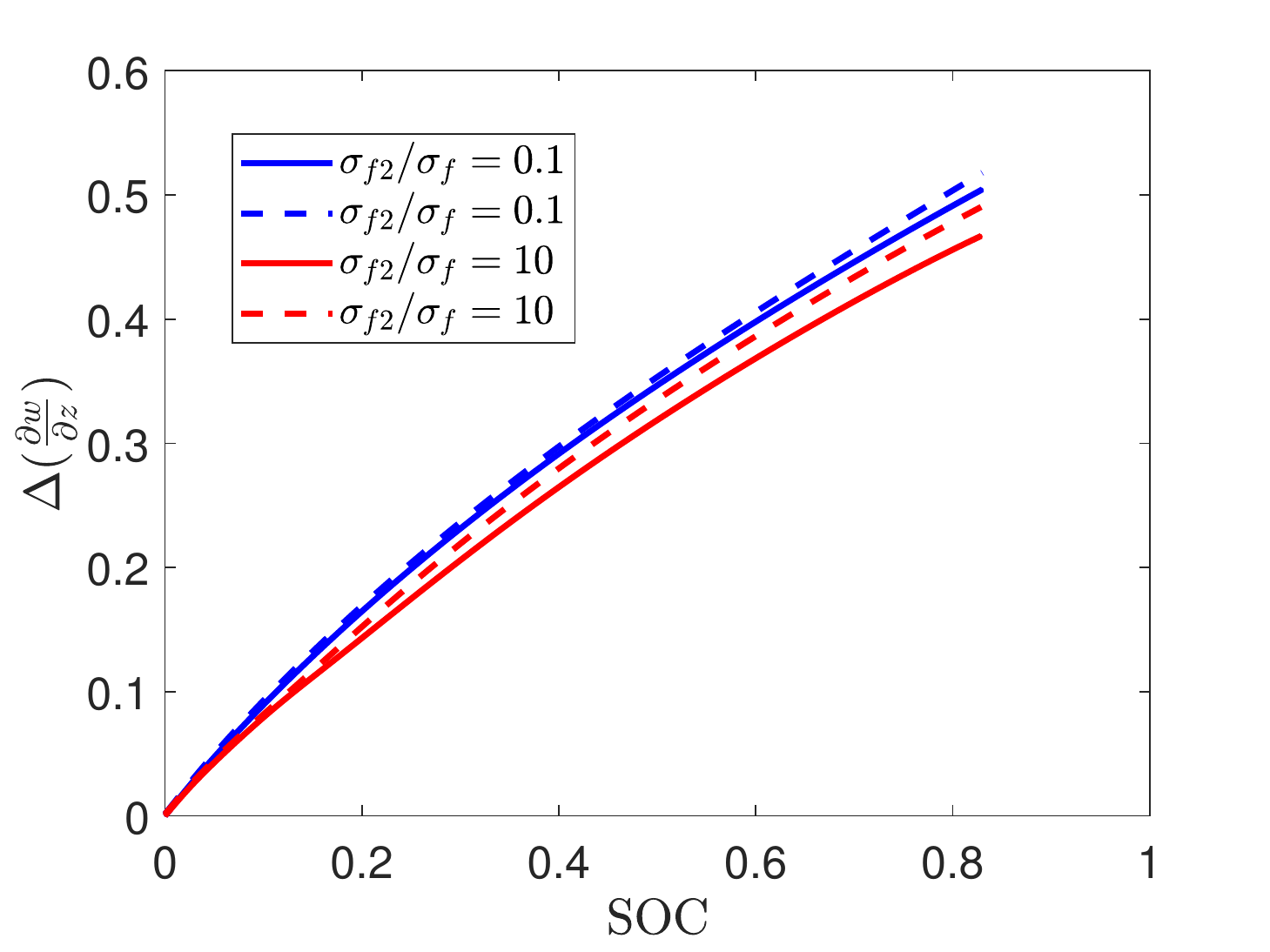}
		\caption{}
	\end{subfigure}
	\caption{(a) Variation of electrode potential with SOC, and (b) variation of length-increase with SOC for two different values of $\frac{\sigma_{f2}}{\sigma_f}$, without (solid lines) and with (dashed lines) considering surface stresses. The charging rate is kept at 20C and the particle size is taken to be 5 nm. } \label{galvanosigmaSSeffect}
\end{figure}

Now that we have delineated the significance of yield strength of the constraining material on the electrode potential and length-increase, we need to understand the effect surface stresses have on both of them. Figure \ref{galvanosigmaSSeffect} shows the effect of surface stresses on the (a) electrode potential and (b) length-increase for two different values of yield-strengths of the constraining material ($\frac{\sigma_{f2}}{\sigma_f}$=0.1, 10) and thickness $r_d=0.3$. As observed in Fig.\ref{SSeffG}(a), the electrode potential shifts downward as a result of surface stresses, the trend is similar here for both $\frac{\sigma_{f2}}{\sigma_f}$ = 0.1 and 10. The magnitude of this shift is not significantly different for the two values of yield-strengths. Coming to Fig. \ref{galvanosigmaSSeffect}(b), we observe an interesting trend. The length-increase is enhanced when surface stress is considered, and the magnitude of this increase in $\frac{\partial w}{\partial z}$ is higher for higher yield strengths of the constraining material. This means, as the ratio of yield-strengths $\frac{\sigma_{f2}}{\sigma_f}$  increases, the effect of surface stress on $\frac{\partial w}{\partial z}$ increases, and it becomes more and more detrimental. Given the compressive effect of surface stress at the outer-periphery, this result is quite counter-intuitive, however a thorough invetigation reveals that when surface stress is considered, the radial expansion of the outer-periphery decreases as compared to without surface stress, as there exists a compressive stress at the outer-periphery in the radial direction. Since the volume-expansion  due to lithiation is similar in both the cases (without and with surface stress), the length increase $\frac{\partial w}{\partial z}$ is enhanced in the latter, to compensate for the reduced radial-expansion. Now, in addition to this, when the yield strength of the constraining material is increased, the radial compressive stress increases at the inner-periphery of the electrode and hence, the radial expansion is further subdued from both the inner and outer peripheries of the electrode particle. As a combined effect, the magnitude of $\frac{\partial w}{\partial z}$ increases further.\\

\begin{figure} [h!]
	\centering
	\begin{subfigure}[b]{0.5\textwidth}
		\includegraphics[width=\textwidth]{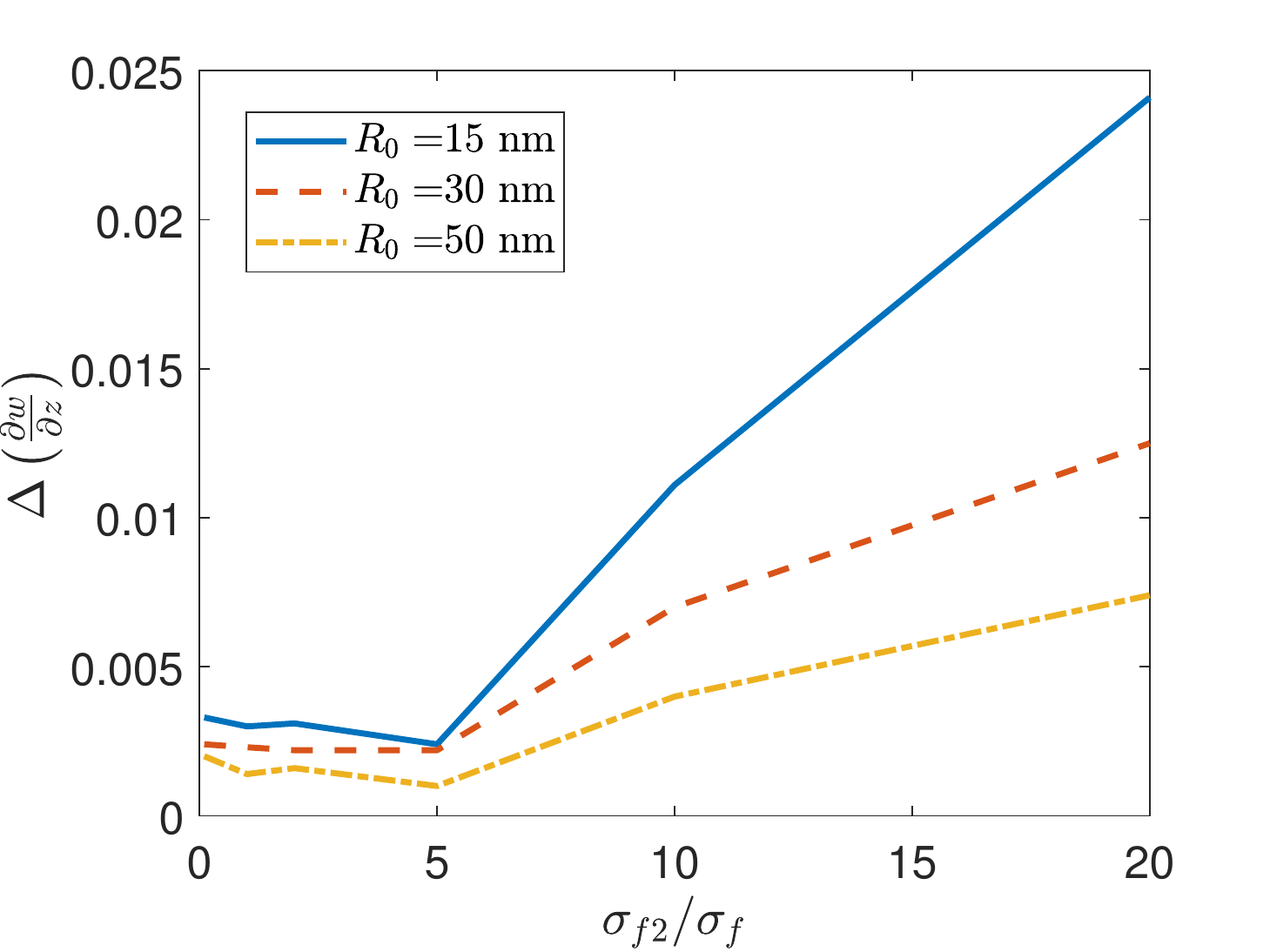}
		\caption{}
	\end{subfigure}%
	\begin{subfigure}[b]{0.5\textwidth}
		\includegraphics[width=\textwidth]{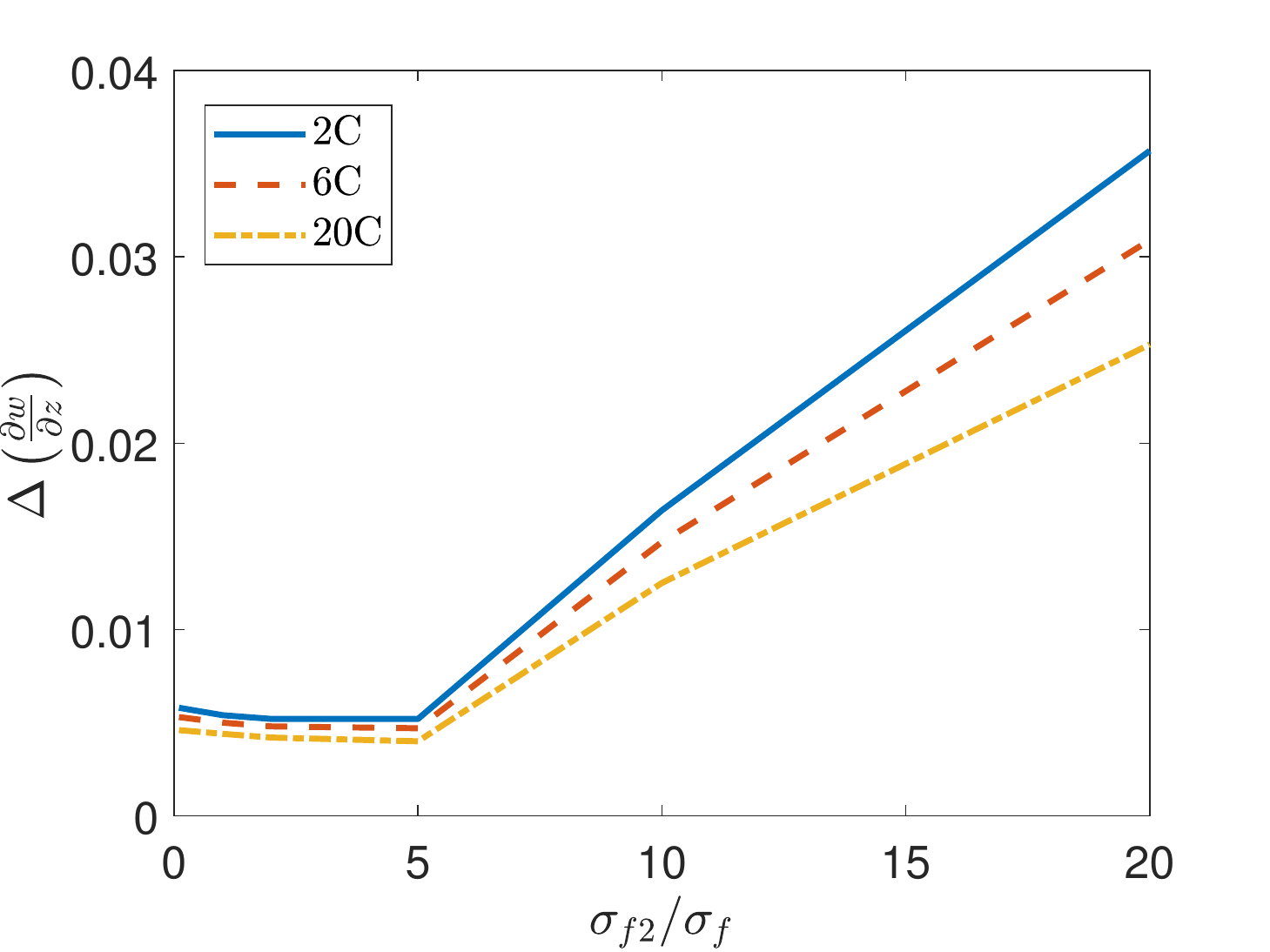}
		\caption{}
	\end{subfigure}
	\caption{(a) Variation of $\Delta\left(\frac{\partial w}{\partial z}\right)$ with $\frac{\sigma_{f2}}{\sigma_f}$ for different particle sizes, keeping charging rate constant at 20C. (b) Variation of $\Delta\left(\frac{\partial w}{\partial z}\right)$ with $\frac{\sigma_{f2}}{\sigma_f}$ for different charging rates, keeping particle size constant at 15 nm.} \label{galvanodelwdash}
\end{figure}

The increase in $\frac{\partial w}{\partial z}$ as an effect of surface stress can be represented by $\Delta(\frac{\partial w}{\partial z})$ which is defined as $\Delta(\frac{\partial w}{\partial z})=(\frac{\partial w}{\partial z})_{\text{w/SS}}-(\frac{\partial w}{\partial z})_{\text{w/o-SS}}$. Figure \ref{galvanodelwdash} depicts the variation of $\Delta(\frac{\partial w}{\partial z})$ with $\frac{\sigma_{f2}}{\sigma_f}$ for different (a) particle sizes, and (a) charging rates. The length-increase is considered at 80$\%$ SOC. From Fig. \ref{galvanodelwdash}(a), it is evident that $\Delta(\frac{\partial w}{\partial z})$ increases with increase in $\frac{\sigma_{f2}}{\sigma_f}$, and with decrease in particle sizes. This result is consistent with a previous work \cite{ASJC2019surfacestress} where surface stress increases in decrease in particle size. On the contrary, in Fig. \ref{galvanodelwdash}(b) we observe a decrease in the magnitude of $\Delta(\frac{\partial w}{\partial z})$ with increase in C-rate. Although this result might seem like contradicting previous results \cite{ASJC2019surfacestress}, but if we look closely at the results obtained in  \cite{chakraborty2015combining}, we would find that the over-all magnitude of length-increase reduces with increase in charging rates. The detailed explanation of this phenomenon exists in the cited document. \\

\subsection{Potentiostatic charging condition} \label{Prnd}

\begin{figure}[h!]
	\centering
	\includegraphics[width=0.6\textwidth]{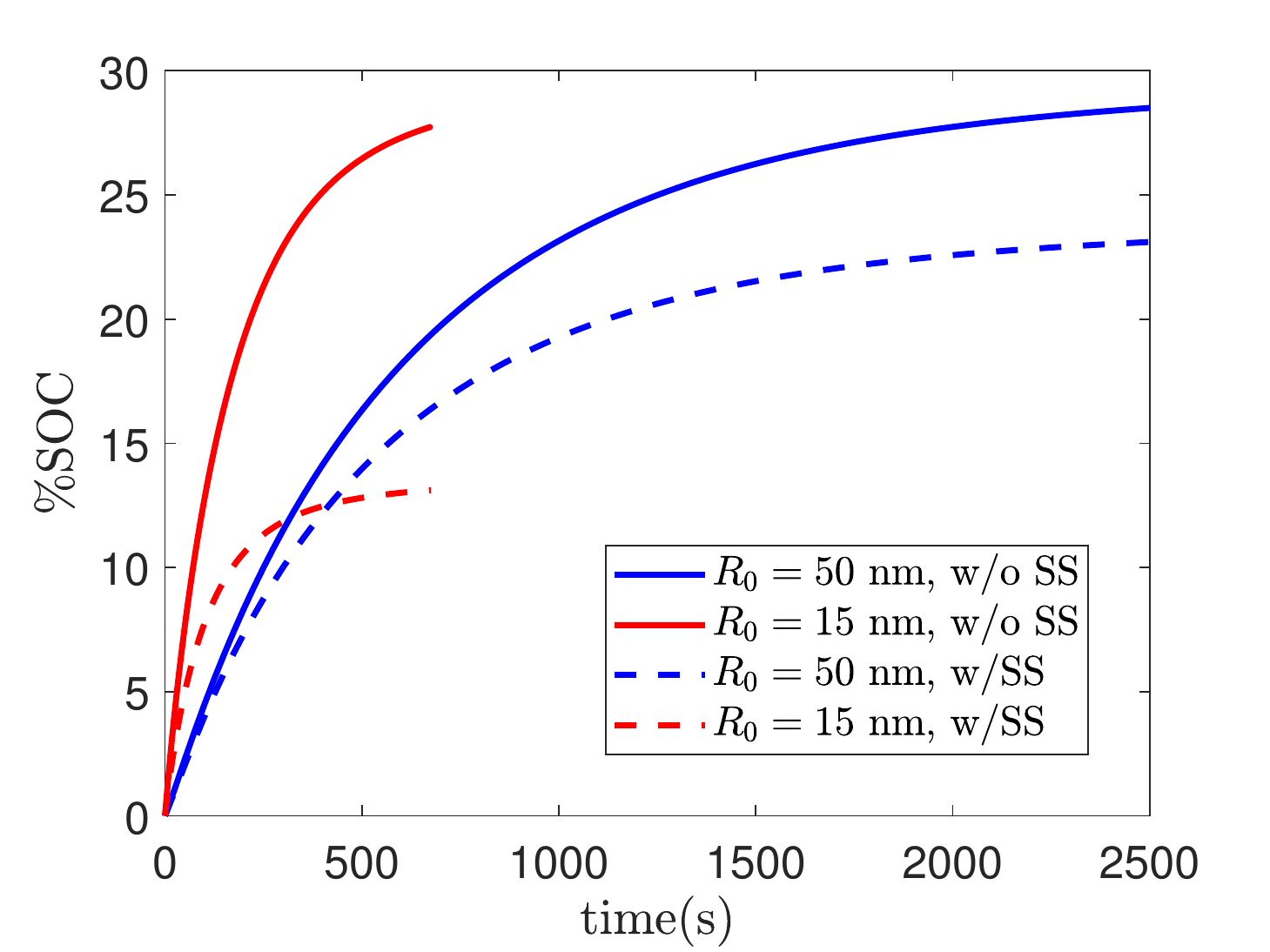}
	\caption{Variation of SOC against time for particles of $R_0 = 50$ and 15 nm, with and without considering surface stress effects.}\label{SSeffectP}
\end{figure}

In potentiostatic condition, as $\Delta \phi_{\rm eq}$ nears $\Delta \phi$, the overpotential value becomes less and less negative. We have set a minimum value for overpotential as $-0.2$ mV \cite{lu2018reaction}. Figure \ref{SSeffectP} shows the SOC attained by particles of varying sizes, with and without surface stress, plotted against dimensionalized time. Since the reduction reaction occurs only when $\eta_o < 0$, the reaction dies down at an overpotential of  $-0.2$ mV, and the SOC reached at this stage is called the equilibrium SOC, because a reaction is at equilibrium when $\eta_o=0$. Here, for the sake of obtaining numerical results we have considered a cut-off value for $\eta_o$ as  $-0.2$ mV. \\ 

\begin{figure} [h!]
	\centering
	\begin{subfigure}[b]{0.35\textwidth}
		\includegraphics[width=\textwidth]{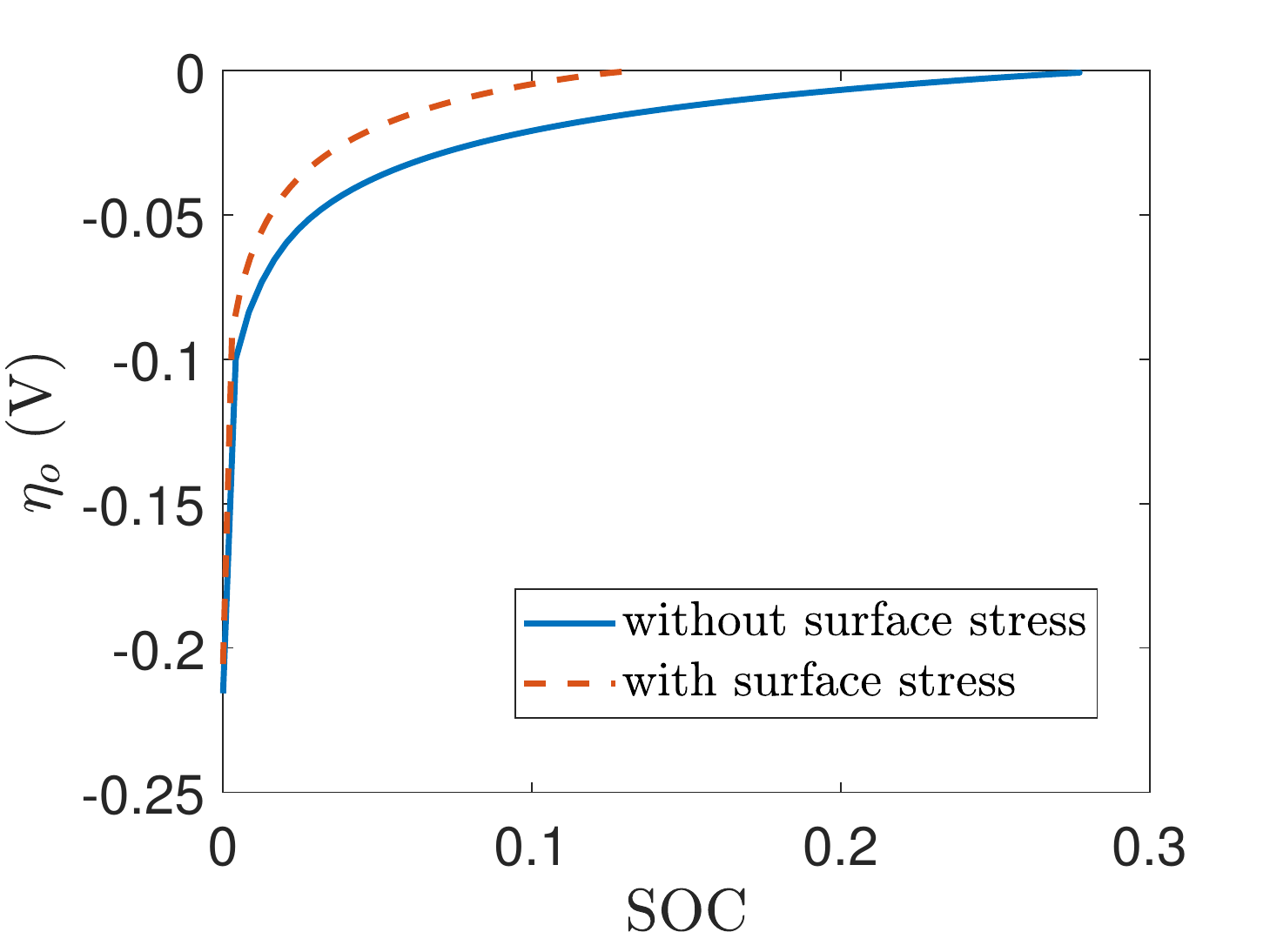}
		\caption{}
	\end{subfigure}%
	\begin{subfigure}[b]{0.35\textwidth}
		\includegraphics[width=\textwidth]{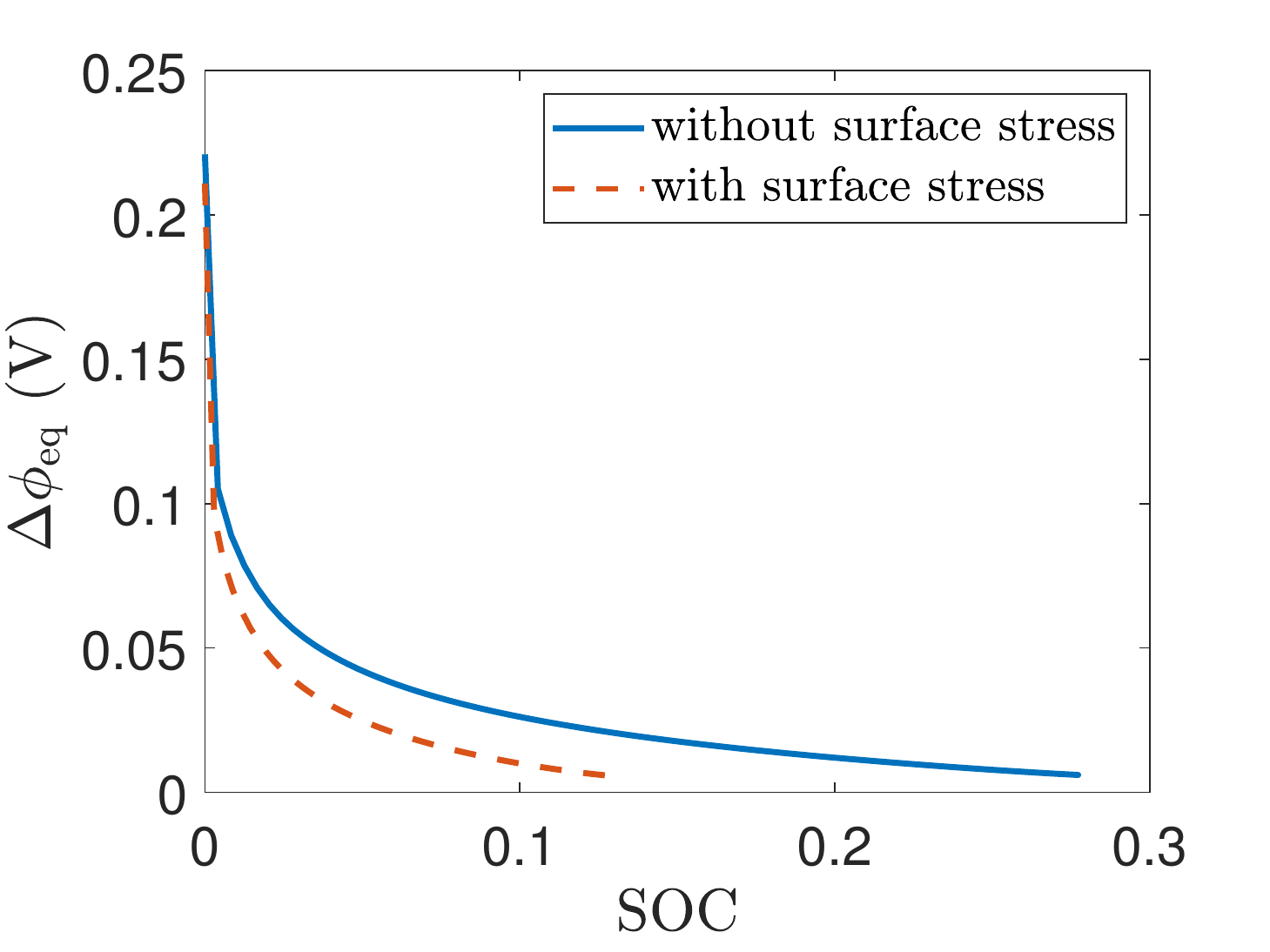}
		\caption{}
	\end{subfigure}%
	\begin{subfigure}[b]{0.35\textwidth}
		\includegraphics[width=\textwidth]{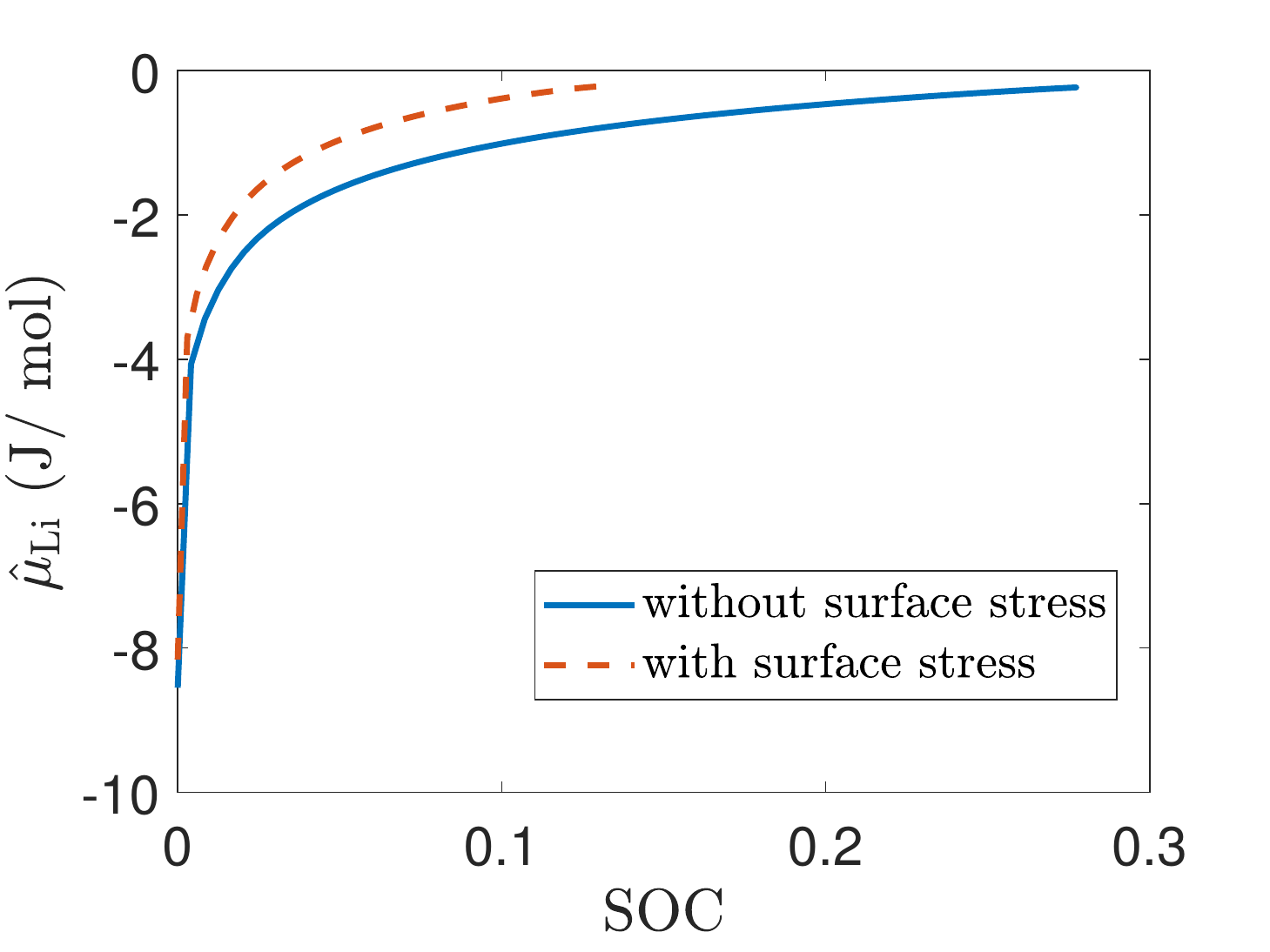}
		\caption{}
	\end{subfigure}
	\caption{Variation of (a) overpotential (b) equilibrium potential, and (c) surface chemical potential, with and without surface stress, plotted against SOC for a potentiostatic charging case in a particle of size 15 nm.} \label{OCE}
\end{figure}

When surface stress is absent, the equilibrium SOC for both the particles (50 and 15 nm) are equal. Intuitively, time taken by Li to diffuse into a larger particle is greater than that for a small particle. So, the reduction reaction at the anode surface and evolution of Li concentration inside the anode particle is independent of particle size. On consideration of surface stress, the equilibrium SOC for both the particles drop to a lower value, and this drop is higher for the smaller particle as surface stress is higher in case of 15 nm particle, as compared to that in 50 nm. Hence, as the charge content within the anode particle decreases when surface stress is considered, we can directly conclude a degradation in the electrochemical performance due to the presence of surface stress. During the course of this study it has come to our notice that the equilibrium SOC is also determined by the form of $\mu_0$ (stress-independent chemical potential) considered, and a detailed discussion of the same is done in the Appendix section.\\

To understand the effects of surface stress better, in Fig. \ref{OCE}, we plot against SOC (a) the overpotential, (b) the equibrium potential, and (c) the chemical potential, with and without surface stress for a particle of initial radius 15 nm. The interface voltage-drop being constant in potentiostatic case, the change in overpotential occurs due to the change in equilibrium potential. Furthermore, the change in equilibrium potential is reflected as a change in the chemical potential and hence, change in the Gibbs free energy at the anode surface where Li ions get reduced to form Li atoms. Similar to the obsevations in case of galvanostatic charging condition (Fig. \ref{SSeffG}), the chemical potential becomes less negative when surface stress is considered, and thus, the spontaneity of Li ions to form Li atoms decreases. This affects the charging and electrochemical performance negatively. 

\begin{figure} [h!]
	\centering
	\begin{subfigure}[b]{0.5\textwidth}
		\includegraphics[width=\textwidth]{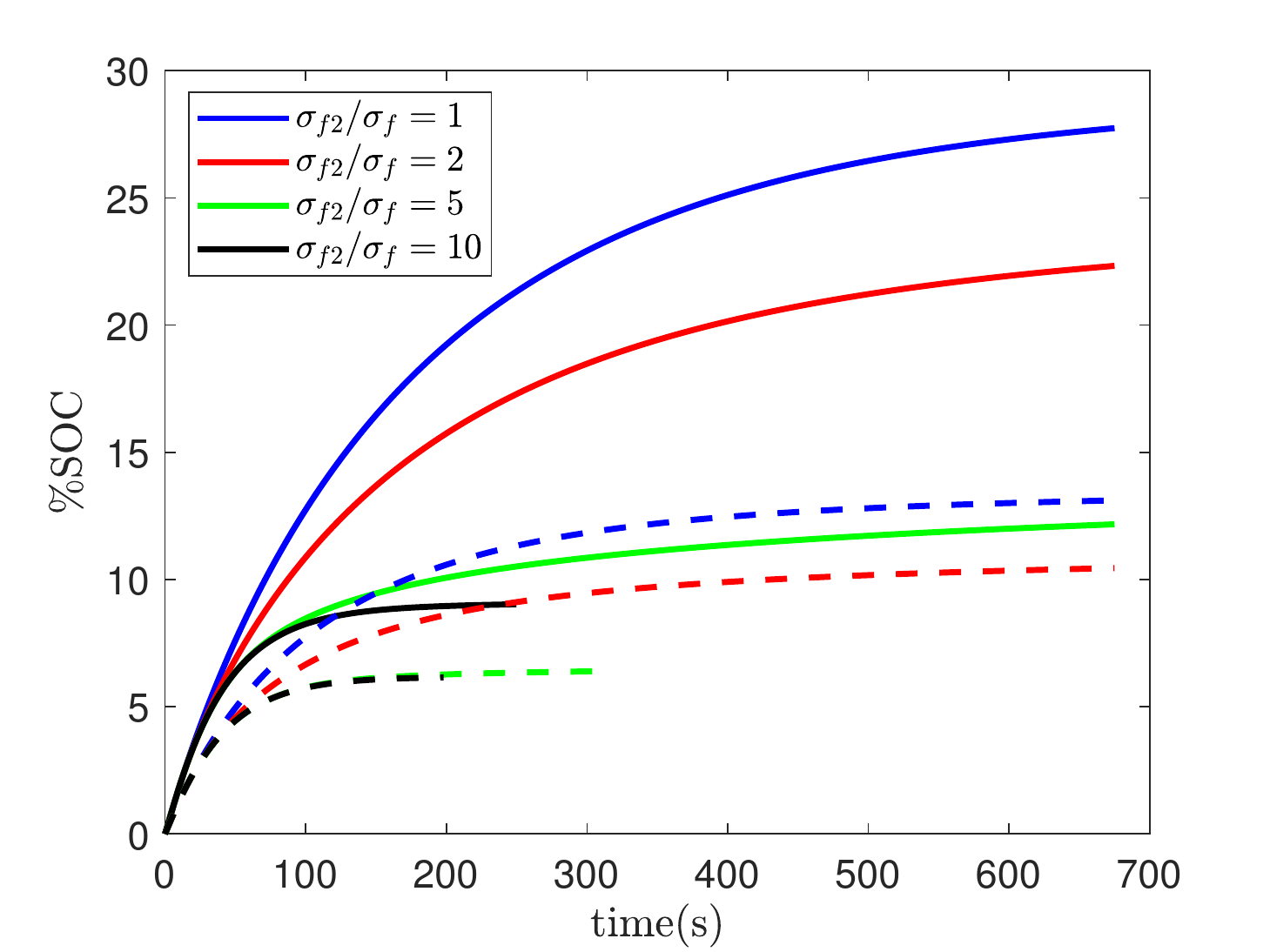}
		\caption{}
	\end{subfigure}%
	\begin{subfigure}[b]{0.5\textwidth}
		\includegraphics[width=\textwidth]{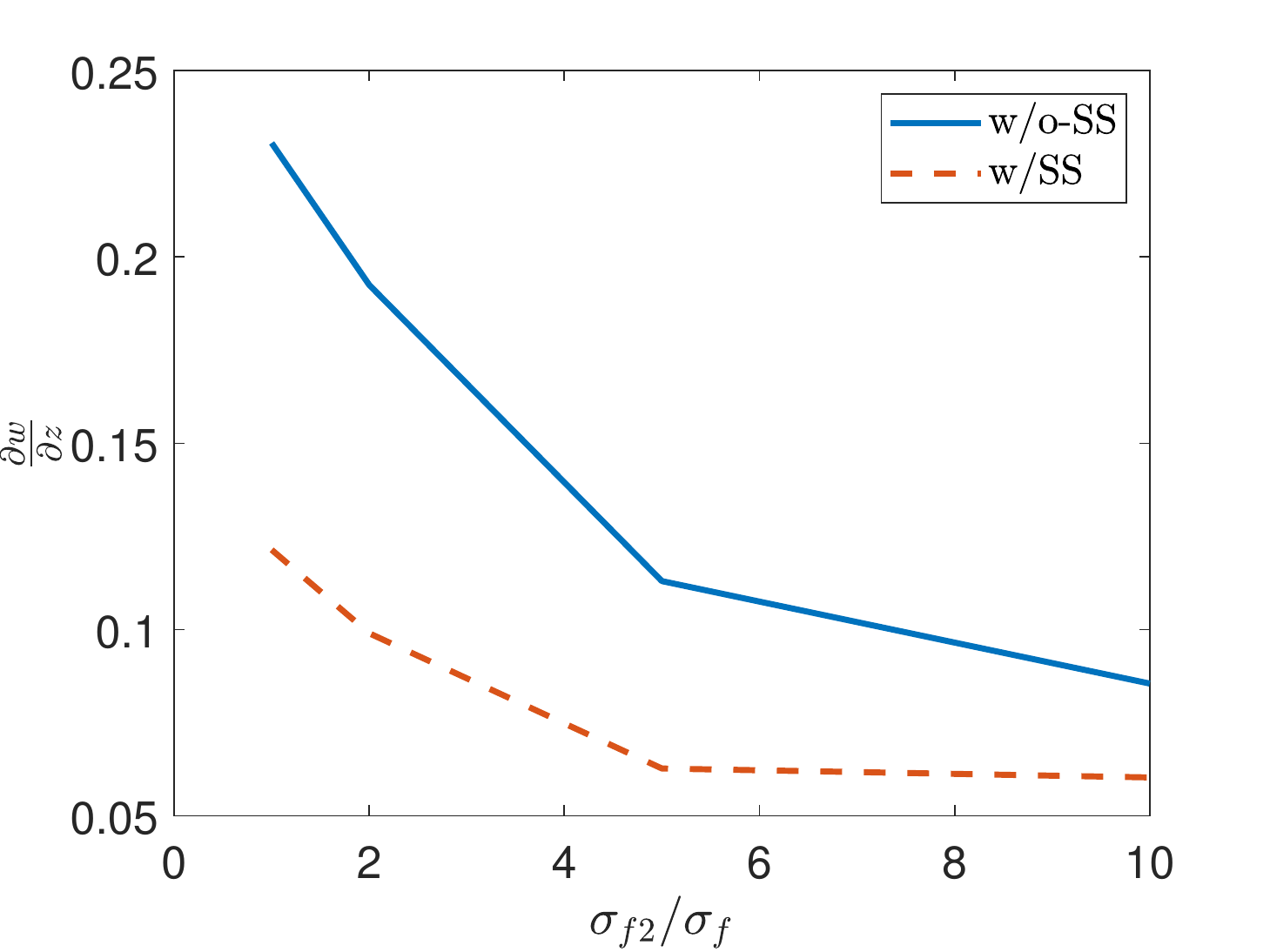}
		\caption{}
	\end{subfigure}
	\caption{(a) Variation of SOC with time for different values of $\frac{\sigma_{f2}}{\sigma_f}$; (b) variation of $\frac{\partial w}{\partial z}$ with $\frac{\sigma_{f2}}{\sigma_f}$ at the equilibrium SOC, without (solid lines) and with (dashed lines) considering surface stresses. The particle size is considered to be 15 nm.} \label{potentioSOCwdash}
\end{figure}

Until now, for the potentiostatic case, the yield strengths of both the constraining material and Si were equal, and we were only studying the effect of surface stress on the electrochemical performance. Just as we proceeded in the galvanostatic case, we would like to see the effects of material properties of the constraining material on the electrochemical performance, as well as mechanical performance (which is demonstrated by the length-increase of the particle, similar to Section 3.1). Figure \ref{potentioSOCwdash}(a) shows the variation of SOC with time, for different values of $\frac{\sigma_{f2}}{\sigma_f}$ (1, 2, 5, 10), without (solid lines) and with (dashed lines) considering the effects of surface stress. The gap between the solid and the dashed lines reduces as the ratio of yield strengths increases. Also, with increase in $\frac{\sigma_{f2}}{\sigma_f}$ value, the equilibrium SOC decreases. For the mechanical performance, Fig.\ref{potentioSOCwdash}(b) shows the variation of $\frac{\partial w}{\partial z}$ at equilibrium SOC with $\frac{\sigma_{f2}}{\sigma_f}$, without (solid line) and with (dashed line) considering surface stress. Quite opposite to what we obtained in galvanostatic charging case, the length-increase magnitudes decrease when surface stress in considered; this occurs because the equilibrium SOC reached with surface stress is lower than that without surface stress (refer Fig.\ref{SSeffectP}) and hence, the overall volumetric expansion of the electrode particle is low due to low lithium concentration. Additionally, the gap between the solid and the dashed lines decreases with increase in $\frac{\sigma_{f2}}{\sigma_f}$. This is consistent with the result in Fig. \ref{potentioSOCwdash}(a), which also sees a reduction in the gap between the solid and dashed lines. 


\subsection{The two competitive outcomes of surface stress} \label{competernd}
\begin{figure} [h!]
	\centering
	\begin{subfigure}[b]{0.5\textwidth}
		\includegraphics[width=\textwidth]{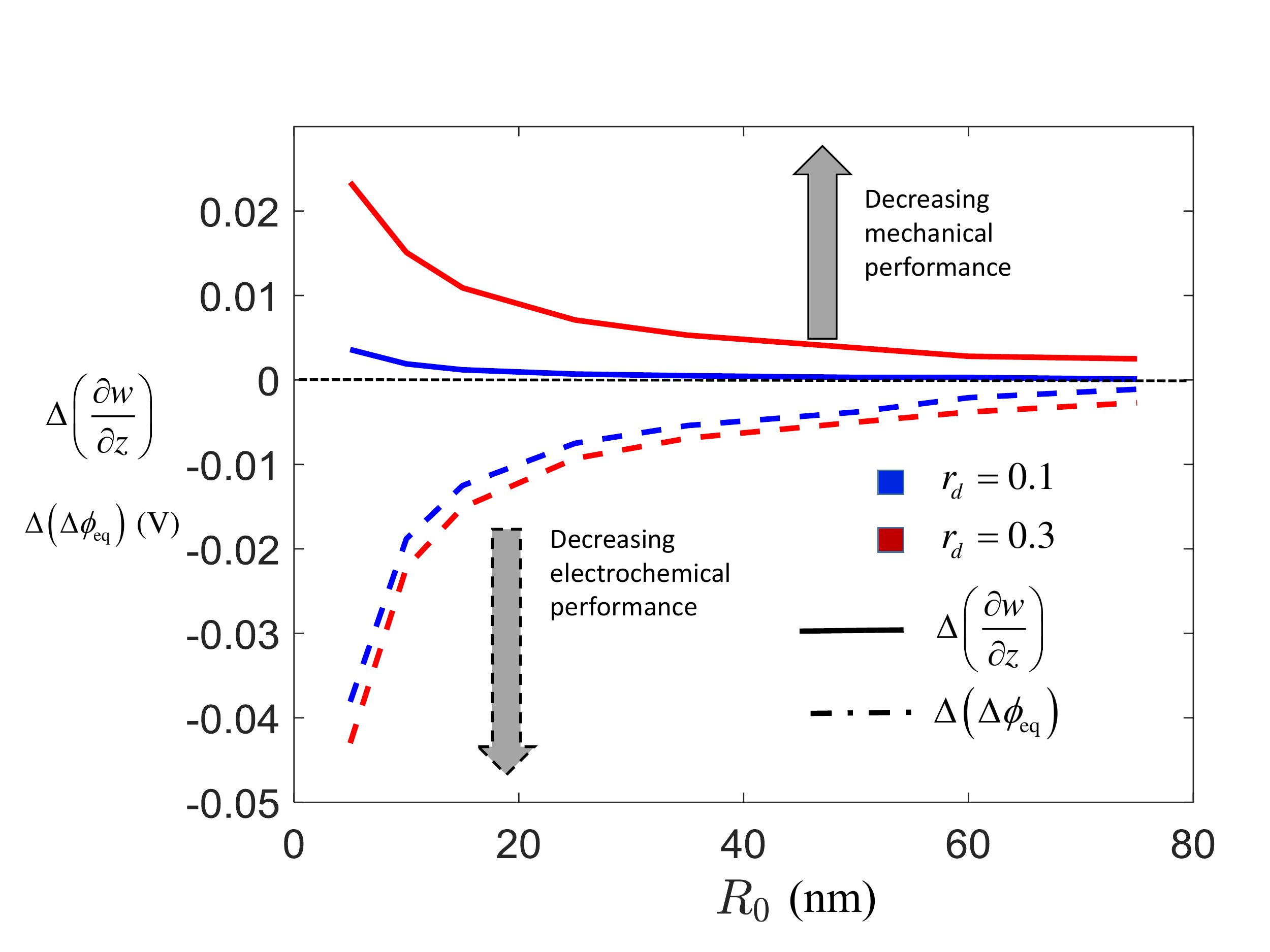}
		\caption{}
	\end{subfigure}%
	\begin{subfigure}[b]{0.5\textwidth}
		\includegraphics[width=\textwidth]{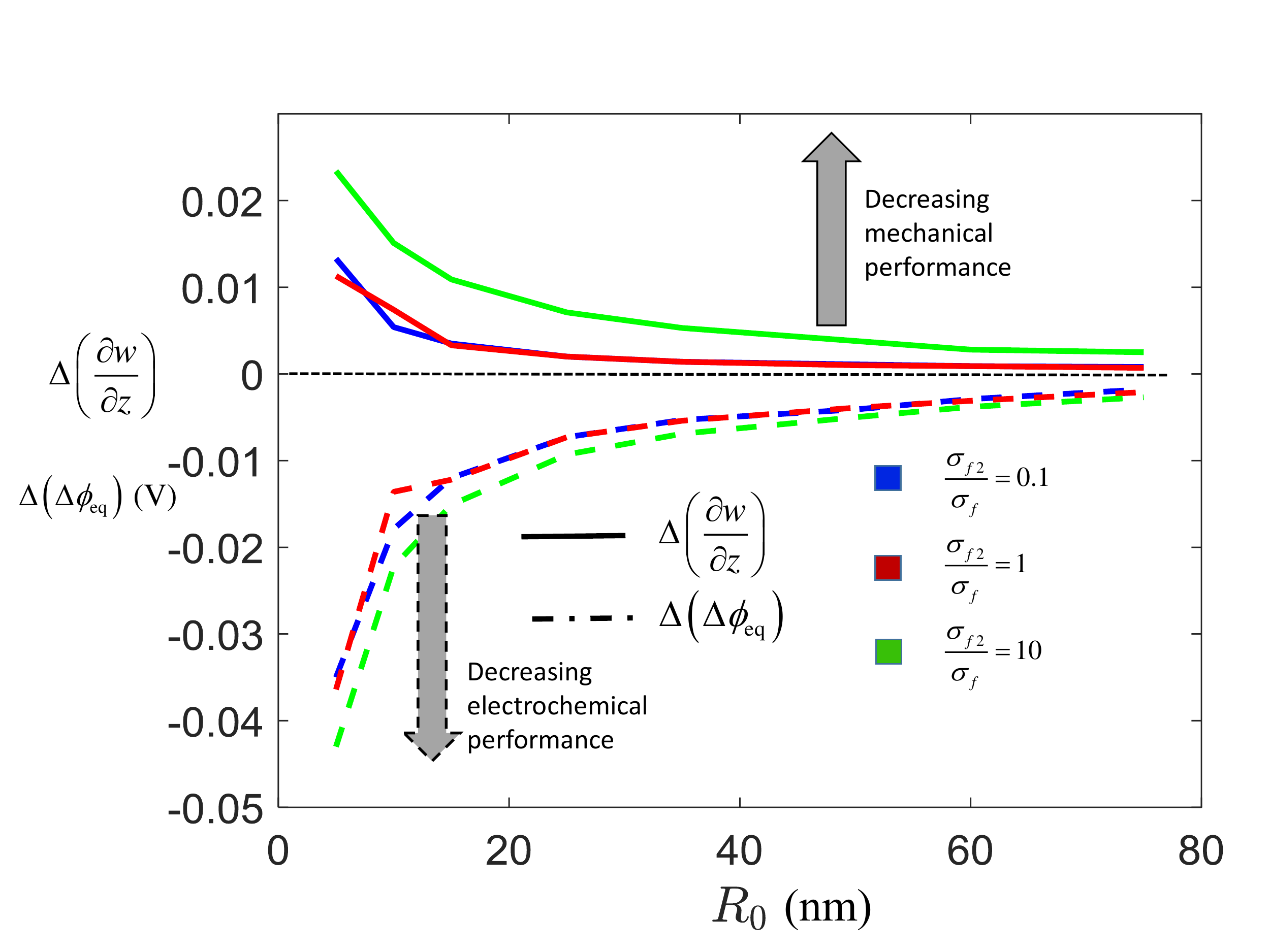}
		\caption{}
	\end{subfigure}
	\begin{subfigure}[b]{0.5\textwidth}
		\includegraphics[width=\textwidth]{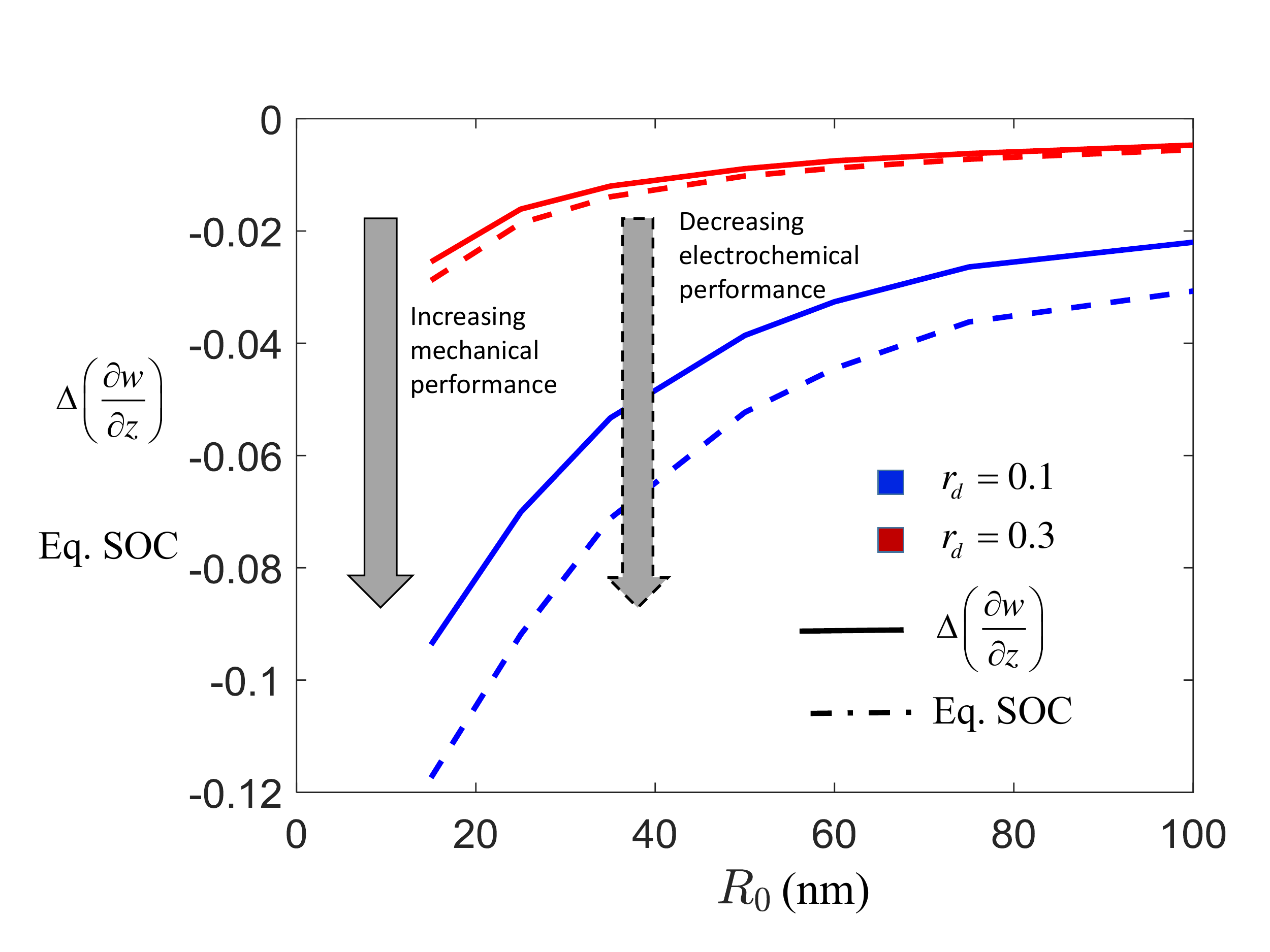}
		\caption{}
	\end{subfigure}%
	\begin{subfigure}[b]{0.5\textwidth}
		\includegraphics[width=\textwidth]{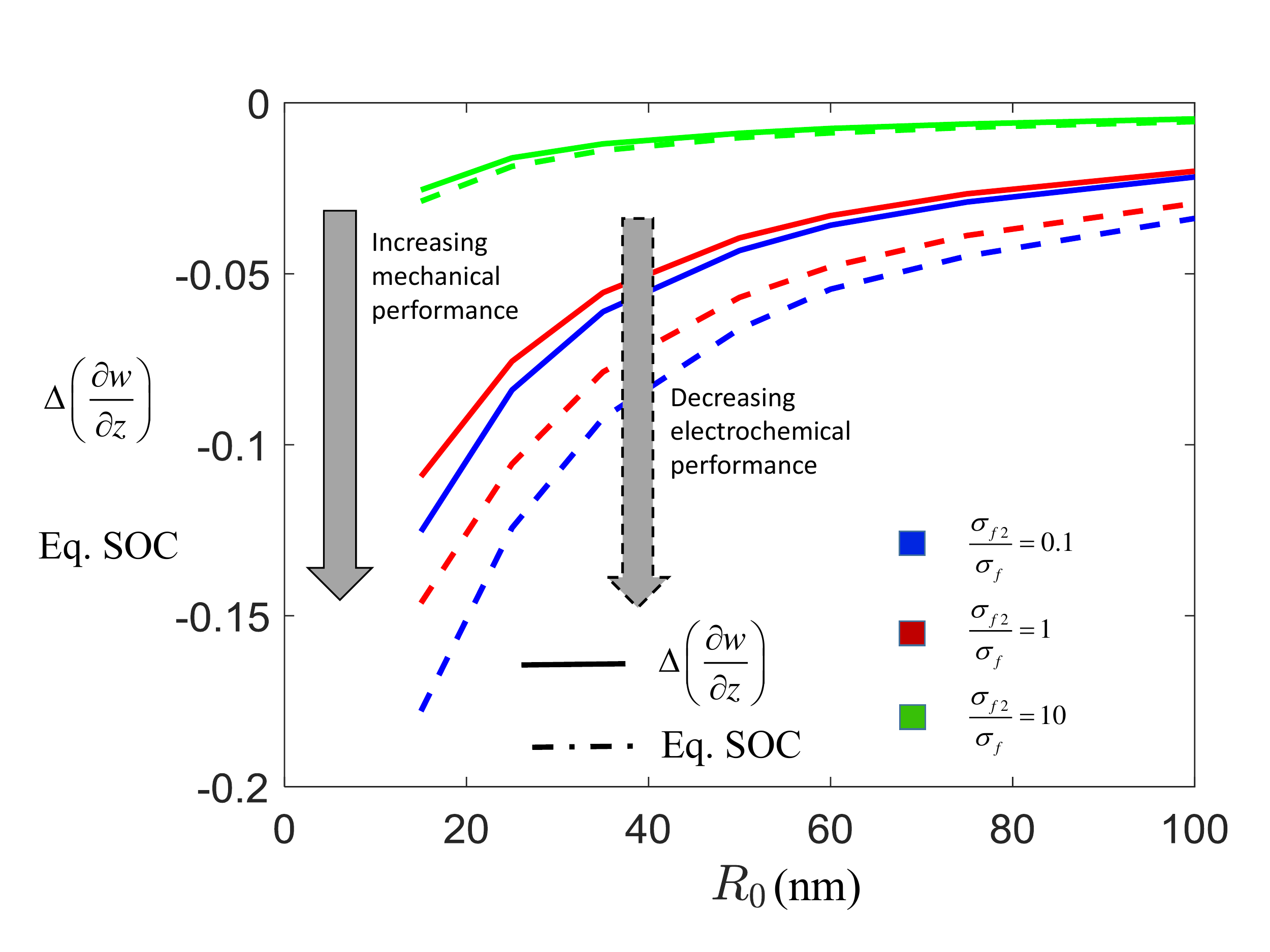}
		\caption{}
	\end{subfigure}
	\caption{Competitive plots for galvanostatic charging condition (a , b) and Potentiostatic charging condition (c ,d). In (a) and (c), core radius of the constraining material $r_d$ is varied (0.1, 0.3) and ($\frac{\sigma_{f2}}{\sigma_f}$) is kept constant at 10. In (b) and (d), the ($\frac{\sigma_{f2}}{\sigma_f}$) is varied (0.1, 1, 10) and $r_d$ is kept constant at 0.3. The solid lines depict mechanical performance, and the dashed lines depict the electrochemical performance. } \label{compete}
\end{figure}

From previous literature  \cite{cheng2008influence,deshpande2010modeling,ASJC2019surfacestress,yang2010insertion,hao2012diffusion} it was established that to make the anode particles safer against mechanical failures in the form of instabilities, fracture, delamination of SEI, and losing contact with the current collector, we should design anode particles with dimensions less than 100 nm. This conclusion was based solely on the fact that surface stresses shift the DISs in the negative direction, hence rendering the particles mechanically safer. In the present model, by considering the length-increase phenomenon, which is another mechanical development, we challenge the previous conclusion, based on two negative outcomes: first, the electrochemical performance of the particles degrades as we reduce their sizes; second, in galvanostatic charging condition, the length-increase of the cylindrical particles increases with increase in surface stress (i.e. decrease in the size of the particles). However, in case of potentiostatic charging condition, owing to low lithium concentration within the particles, the length-increase magnitude decreases as we reduce the particle-size. Now, when we add the influence of a constraining material at the inner core, the over-all effects of the surface stress and constraining material on the mechanical and electrochemical performance of the electrode particles needs to be investigated, in tandem. The effects of the constraining material could be analyzed by varying the physical properties (for eg. the radial thickness of the constraint), or the material properties (for eg. the yield strength of the material). In order to arrive at a competitive analysis, we plot a parameter each representing mechanical and electrochemical aspects, under the combined effects of physical/material properties of the constraining material and surface stress (by varying the particle size). 

\subsubsection{Galvanostatic charging condition}
Figures \ref{compete}(a) and (b) are for the case of galvanostatic charging. The difference $\Delta(\text{quantity})$ is defined as $(\text{quantity})_{\text{with surface stress}}-(\text{quantity})_{\text{without surface stress}}$. Therefore, as the magnitude of $\Delta$ increases, it would imply that the effect of surface stress increases.  For Fig.\ref{compete}(a) we vary the core radius $r_d$, keeping the yield strength ratio ($\frac{\sigma_{f2}}{\sigma_f}$) constant at 10. It is important to note that the magnitude of $\Delta(\frac{\partial w}{\partial z})$ is positive (discussed earlier for Fig. \ref{galvanosigmaSSeffect}), which means the mechanical performance degrades in the presence of surface stress. Similarly, the magnitude of $\Delta(\Delta \phi_{\rm eq})$ is negative, which means the electrochemical performance degrades when surface stress is considered. The particle size is varied from 5 to 100 nm. As the particle size decreases, the magnitude of $Delta$ increases, for both length-increase ($\frac{\partial w}{\partial z}$) and equilibrium potential ($\Delta \phi_{\rm eq}$). Hence, with increase in surface stresses, both the mechanical and electrochemical performances degrade. This decrease in performances is further aided by an increase in the core radius of the constraining material. 

Next, if we keep the core radius constant and change the yield strength of constraining material instead, we observe that with increase in the yield strength of the constraining material as compared to that of Si, the effect of surface stresses increase, \textit{i.e.} the magnitude of $\Delta$ increases, thus decreasing both the mechanical and electrochemical performances. 

\subsubsection{Potentiostatic charging condition}
Figures \ref{compete}(c) and (d)
 are for the case of potentiostatic charging condition. Contrary to our observations for the galvanostatic cases, here we observe a negative value for the magnitude of $\Delta(\frac{\partial w}{\partial z})$ , which means the mechanical performance improves with the advent of surface stress. However, the electrochemical performance degrades here as well. The particle size is varied from 15 to 100 nm. As the particle size decreases, the surface stresses become more prominent, which leads to an increased mechanical performance and a reduced electrochemical performance. The effect of core radius on surface stress is opposite to what we had observed in the galvanostatic case. Here, with increase in core radius of the constraining material, the effect of surface stress decreases.
 
 When we keep the core radius constant, and vary the yield strength ratio ($\frac{\sigma_{f2}}{\sigma_f}$), again we observe an opposite effect from that of galvanostatic case. As $\frac{\sigma_{f2}}{\sigma_f}$ increases, the effect of surface stress decreases. 


\section{Conclusion} \label{conclusion}
We developed a fully coupled electro-chemo-mechanical formulation for a nano-structured, cylindrical silicon anode particle undergoing finite deformation due to lithiation/de-lithiation to study the effects of surface stress on the electrochemical and mechanical performances of the anode particles. We analyze the effects of increasing surface stress under two different charging/discharging conditions: first, galvanostatic case with constant charging/discharging rate; second, potentiostatic case with constant interface voltage drop. 

As observed by previous researchers  \cite{cheng2008influence, deshpande2010modeling, yang2010insertion, hao2012diffusion, ASJC2019surfacestress}, the compressive effect of surface stress relaxes the diffusion-induced stresses within the anode particle, thus increasing their structural integrity. However, when we considered the phenomenon of length increase of the particles upon lithiation, and how surface stress affects it, we obtain an interesting result based on the kind of charging/discharging condition. Most importantly, it has been observed that the growing surface stress magnitudes can have a detrimental effect on the electrochemical performance of the battery. It is important to highlight the four major observations:

\begin{itemize}
\item As surface stress gets significant, the surface energy of the anode particles increases. During charging, as lithium ions reduce to form lithium atoms at the surface of the anode particles, the chemical potential at the Si surface becomes less negative. This implies, that the electrochemical reaction becomes less spontaneous in the presence of surface stress. This manifests itself (a) by a negative shift of the interfacial voltage-drop curve in case of galvanostatic charging, and (b) by reaching a lower equilibrium SOC in case of potentiostatic charging, when compared to without surface stress case. 

\item As we decrease the size of the particle, the magnitude of the surface stress increases. Therefore, with a decreasing particle size, the negative shift in interfacial voltage-drop increases for galvanostatic case, and the reduction in equilibrium SOC increases for potentiostatic case. 

\item With increase in magnitude of surface stress for a decreasing particle size, the magnitude of length-increase of the particles enhances for a galvanostatic case, and reduces for a potentiostatic case. However, as observed from previous literature, in both the cases, the compressive effect of surface stress relaxes the diffusion-induced stresses within the anode particle, apparently giving an impression of improvement in structural integrity being directly proportional to decreasing particle-size. This straight-forward inference is challenged through the results in the present work. In galvanostatic case, an unrestricted reduction in particle-size may lead to an overwhelmed length-increase, which has negative consequences on the mechanical performance of the battery. 

\item The introduction of a constraining material at the core of the Si nanotube makes the investigation of the length-increase phenomenon complete. As we increase the yield-strength or/and the core radius of the constraining material (neglecting surface stress), the length-increase magnitude decreases. 	
\end{itemize}

Finally, the two competitive phenomena of increasing mechanical safety and degrading electrochemical performance can be compared to choose the best possible particle sizes for a potentiostatic case. Whereas, for a galvanostatic case, the particle sizes should not be made significantly small. 

\section{Acknowledgement}
AS and JC acknowledge the financial support of Ministry of Human Resource Development, Government of India. JC thanks the DST-INSPIRE program of the Government of India (DST/INSPIRE Faculty Award/2016/DST/INSPIRE/04/2015/002825) as well as the ISIRD Funding (IIT/SRIC/ME/MFD/2017-18/70) from IIT Kharagpur for the computational resources.

\clearpage
\bibliography{bibdocument}

\begin{thebibliography}{10}

\bibitem{mizushima1981lixcoo2}
K~Mizushima, PC~Jones, PJ~Wiseman, and John~B Goodenough.
\newblock Lixcoo2 (0< x⩽ 1): A new cathode material for batteries of high
  energy density.
\newblock {\em Solid State Ionics}, 3:171--174, 1981.

\bibitem{whittingham1979intercalation}
M~Stanley Whittingham.
\newblock Intercalation chemistry and energy storage.
\newblock {\em Journal of Solid State Chemistry}, 29(3):303--310, 1979.

\bibitem{yoshino2014development}
Akira Yoshino.
\newblock Development of the lithium-ion battery and recent technological
  trends.
\newblock In {\em Lithium-ion batteries}, pages 1--20. Elsevier, 2014.

\bibitem{huggins1999lithium}
Robert~A Huggins.
\newblock Lithium alloy negative electrodes.
\newblock {\em Journal of Power Sources}, 81:13--19, 1999.

\bibitem{wu2018modeling}
Hui Wu, Zhoucan Xie, Yan Wang, Chunsheng Lu, and Zengsheng Ma.
\newblock Modeling diffusion--induced stress on two-phase lithiation in
  lithium-ion batteries.
\newblock {\em European Journal of Mechanics-A/Solids}, 71:320--325, 2018.

\bibitem{ozanam2016silicon}
Fran{\c{c}}ois Ozanam and Michel Rosso.
\newblock Silicon as anode material for li-ion batteries.
\newblock {\em Materials Science and Engineering: B}, 213:2--11, 2016.

\bibitem{beaulieu2001colossal}
LY~Beaulieu, KW~Eberman, RL~Turner, LJ~Krause, and JR~Dahn.
\newblock Colossal reversible volume changes in lithium alloys.
\newblock {\em Electrochemical and Solid-State Letters}, 4(9):A137--A140, 2001.

\bibitem{wang1999tem}
Haifeng Wang, Young-Il Jang, Biying Huang, Donald~R Sadoway, and Yet-Ming
  Chiang.
\newblock {TEM} {S}tudy of {E}lectrochemical {C}ycling-{I}nduced {D}amage and
  {D}isorder in {L}i{C}o{O}2 {C}athodes for {R}echargeable {L}ithium
  {B}atteries.
\newblock {\em Journal of The Electrochemical Society}, 146(2):473--480, 1999.

\bibitem{kasavajjula2007nano}
Uday Kasavajjula, Chunsheng Wang, and A~John Appleby.
\newblock Nano-and bulk-silicon-based insertion anodes for lithium-ion
  secondary cells.
\newblock {\em Journal of Power Sources}, 163(2):1003--1039, 2007.

\bibitem{ko2015challenges}
Minseong Ko, Sujong Chae, and Jaephil Cho.
\newblock Challenges in accommodating volume change of si anodes for li-ion
  batteries.
\newblock {\em ChemElectroChem}, 2(11):1645--1651, 2015.

\bibitem{tarascon2001issues}
J-M Tarascon and M~Armand.
\newblock Issues and challenges facing rechargeable lithium batteries:
  {M}aterials for clean energy.
\newblock {\em Nature}, 414(6861):359--367, 2001.

\bibitem{muller2018quantification}
Simon M{\"u}ller, Patrick Pietsch, Ben-Elias Brandt, Paul Baade, Vincent
  De~Andrade, Francesco De~Carlo, and Vanessa Wood.
\newblock Quantification and modeling of mechanical degradation in lithium-ion
  batteries based on nanoscale imaging.
\newblock {\em Nature communications}, 9(1):2340, 2018.

\bibitem{ebner2013visualization}
Martin Ebner, Federica Marone, Marco Stampanoni, and Vanessa Wood.
\newblock Visualization and quantification of electrochemical and mechanical
  degradation in {L}i ion batteries.
\newblock {\em Science}, 342(6159):716--720, 2013.

\bibitem{kabir2017degradation}
MM~Kabir and Dervis~Emre Demirocak.
\newblock Degradation mechanisms in li-ion batteries: a state-of-the-art
  review.
\newblock {\em International Journal of Energy Research}, 41(14):1963--1986,
  2017.

\bibitem{reniers2019review}
Jorn~M Reniers, Grietus Mulder, and David~A Howey.
\newblock Review and performance comparison of mechanical-chemical degradation
  models for lithium-ion batteries.
\newblock {\em Journal of The Electrochemical Society}, 166(14):A3189, 2019.

\bibitem{liu2020safety}
Binghe Liu, Yikai Jia, Chunhao Yuan, Lubing Wang, Xiang Gao, Sha Yin, and Jun
  Xu.
\newblock Safety issues and mechanisms of lithium-ion battery cell upon
  mechanical abusive loading: A review.
\newblock {\em Energy Storage Materials}, 24:85--112, 2020.

\bibitem{li2020comprehensive}
J~Li, Robert~G Landers, and Jonghyun Park.
\newblock A comprehensive single-particle-degradation model for battery
  state-of-health prediction.
\newblock {\em Journal of Power Sources}, 456:227950, 2020.

\bibitem{srivastav2017modelling}
Shruti Srivastav, Chao Xu, Kristina Edstr{\"o}m, Torbj{\"o}rn Gustafsson, and
  Daniel Brandell.
\newblock Modelling the morphological background to capacity fade in si-based
  lithium-ion batteries.
\newblock {\em Electrochimica Acta}, 258:755--763, 2017.

\bibitem{xu2015electrode}
Jiagang Xu, Rutooj~D Deshpande, Jie Pan, Yang-Tse Cheng, and Vincent~S
  Battaglia.
\newblock Electrode side reactions, capacity loss and mechanical degradation in
  lithium-ion batteries.
\newblock {\em Journal of The Electrochemical Society}, 162(10):A2026--A2035,
  2015.

\bibitem{chakraborty2015influence}
Jeevanjyoti Chakraborty, Colin~P Please, Alain Goriely, and S~Jonathan Chapman.
\newblock Influence of constraints on axial growth reduction of cylindrical
  li-ion battery electrode particles.
\newblock {\em Journal of Power Sources}, 279:746--758, 2015.

\bibitem{2016Bailinbuckling}
Kai Zhang, Yong Li, Bailin Zheng, Gangpeng Wu, Jingshen Wu, and Fuqian Yang.
\newblock Large deformation analysis of diffusion-induced buckling of nanowires
  in lithium-ion batteries.
\newblock {\em International Journal of Solids and Structures}, 108, 12 2016.

\bibitem{2018BailinZheng}
Kai Zhang, Yong Li, Jingshen Wu, Bailin Zheng, and Fuqian Yang.
\newblock Lithiation-induced buckling of wire-based electrodes in lithium-ion
  batteries: A phase-field model coupled with large deformation.
\newblock {\em International Journal of Solids and Structures}, 144-145, 05
  2018.

\bibitem{zhang2018lithiation}
Kai Zhang, Yong Li, Jingshen Wu, Bailin Zheng, and Fuqian Yang.
\newblock Lithiation-induced buckling of wire-based electrodes in lithium-ion
  batteries: A phase-field model coupled with large deformation.
\newblock {\em International Journal of Solids and Structures}, 144:289--300,
  2018.

\bibitem{zhang85buckling}
Yuwei Zhang, Siyuan Zhan, Kai Zhang, Bailin Zheng, and Liangxinbu Lyu.
\newblock Buckling behavior of a wire-like electrode with a
  concentration-dependent elastic modulus based on a deformed configuration.
\newblock {\em European Journal of Mechanics-A/Solids}, 85:104111.

\bibitem{miehe2016phase}
Christian Miehe, H~Dal, L-M Sch{\"a}nzel, and A~Raina.
\newblock A phase-field model for chemo-mechanical induced fracture in
  lithium-ion battery electrode particles.
\newblock {\em International Journal for Numerical Methods in Engineering},
  106(9):683--711, 2016.

\bibitem{xu2019analytical}
Chengjun Xu, Li~Weng, Lian Ji, and Jianqiu Zhou.
\newblock An analytical model for the fracture behavior of the flexible
  lithium-ion batteries under bending deformation.
\newblock {\em European Journal of Mechanics-A/Solids}, 73:47--56, 2019.

\bibitem{perassi2019capacity}
Eduardo~M Perassi and Ezequiel~PM Leiva.
\newblock Capacity fading model for a solid electrolyte interface with surface
  growth.
\newblock {\em Electrochimica Acta}, 2019.

\bibitem{swaminathan2007electrochemomechanical}
N~Swaminathan, J~Qu, and Y~Sun.
\newblock An electrochemomechanical theory of defects in ionic solids. i.
  theory.
\newblock {\em Philosophical Magazine}, 87(11):1705--1721, 2007.

\bibitem{swaminathan2007electrochemomechanical2}
N~Swaminathan, J~Qu, and Y~Sun.
\newblock An electrochemomechanical theory of defects in ionic solids. part ii.
  examples.
\newblock {\em Philosophical Magazine}, 87(11):1723--1742, 2007.

\bibitem{iqbal2020chemo}
Noman Iqbal, Yasir Ali, and Seungjun Lee.
\newblock Chemo-mechanical response of composite electrode systems with
  multiple binder connections.
\newblock {\em Electrochimica Acta}, 364:137312, 2020.

\bibitem{ram2020JAP}
Ram~Hemanth Yeerella, Hemanth Sai~Sandeep Boddeda, Amrita Sengupta, and
  Jeevanjyoti Chakraborty.
\newblock Role of in situ electrode environments in mitigating
  instability-induced battery degradation.
\newblock {\em Journal of Applied Physics}, 128(23):234901, 2020.

\bibitem{pan2020effect}
Zhexin Pan, Tobias Sedlatschek, and Yong Xia.
\newblock Effect of state-of-charge and air exposure on tensile mechanical
  properties of lithium-ion battery electrodes.
\newblock {\em Journal of The Electrochemical Society}, 167(9):090517, 2020.

\bibitem{liu2020cracks}
Binghe Liu and Jun Xu.
\newblock Cracks of silicon nanoparticles in anodes:
  Mechanics--electrochemical-coupled modeling framework based on the
  phase-field method.
\newblock {\em ACS Applied Energy Materials}, 2020.

\bibitem{chakraborty2015combining}
Jeevanjyoti Chakraborty, Colin~P Please, Alain Goriely, and S~Jonathan Chapman.
\newblock Combining mechanical and chemical effects in the deformation and
  failure of a cylindrical electrode particle in a {L}i-ion battery.
\newblock {\em International Journal of Solids and Structures}, 54:66--81,
  2015.

\bibitem{arico2005nanostructured}
Antonino~Salvatore Arico, Peter Bruce, Bruno Scrosati, Jean-Marie Tarascon, and
  Walter Van~Schalkwijk.
\newblock Nanostructured materials for advanced energy conversion and storage
  devices.
\newblock {\em Nature Materials}, 4(5):366, 2005.

\bibitem{lewis2007situ}
RB~Lewis, A~Timmons, RE~Mar, and JR~Dahn.
\newblock In situ {AFM} measurements of the expansion and contraction of
  amorphous {S}n-{C}o-{C} films reacting with lithium.
\newblock {\em Journal of The Electrochemical Society}, 154(3):A213--A216,
  2007.

\bibitem{chan2008high}
Candace~K Chan, Hailin Peng, Gao Liu, Kevin McIlwrath, Xiao~Feng Zhang,
  Robert~A Huggins, and Yi~Cui.
\newblock High-performance lithium battery anodes using silicon nanowires.
\newblock {\em Nature nanotechnology}, 3(1):31, 2008.

\bibitem{graetz2003highly}
Jason Graetz, CC~Ahn, Rachid Yazami, and Brent Fultz.
\newblock Highly reversible lithium storage in nanostructured silicon.
\newblock {\em Electrochemical and Solid-State Letters}, 6(9):A194--A197, 2003.

\bibitem{xie2015phase}
Yuanyuan Xie, Ming Qiu, Xianfeng Gao, Dongsheng Guan, and Chris Yuan.
\newblock Phase field modeling of silicon nanowire based lithium ion battery
  composite electrode.
\newblock {\em Electrochimica Acta}, 186:542--551, 2015.

\bibitem{lai2019silicon}
Samson~Yuxiu Lai, Thomas~J Preston, Marte~O Skare, Hallgeir Klette,
  Kenneth~Dahl Knudsen, Jan~Petter M{\ae}hlen, and Alexey~Y Koposov.
\newblock Silicon nanoparticles: Size and morphology effects in lithium ion
  batteries.
\newblock In {\em Meeting Abstracts}, number~2, pages 176--176. The
  Electrochemical Society, 2019.

\bibitem{srijan2018JAP}
Srijan Neogi and Jeevanjyoti Chakraborty.
\newblock Size-dependent effects sensitively determine buckling of a
  cylindrical silicon electrode particle in a lithium-ion battery.
\newblock {\em Journal of Applied Physics}, 124(15):154302, 2018.

\bibitem{nasr2019surface}
Mohammad Nasr~Esfahani.
\newblock Surface stress effects on the mechanical properties of silicon
  nanowires: A molecular dynamics simulation.
\newblock {\em Journal of Applied Physics}, 125(13):135101, 2019.

\bibitem{chan2011high}
Candace~K Chan, Hailin Peng, Gao Liu, Kevin McIlwrath, Xiao~Feng Zhang,
  Robert~A Huggins, and Yi~Cui.
\newblock High-performance lithium battery anodes using silicon nanowires.
\newblock In {\em Materials for Sustainable Energy: A Collection of
  Peer-Reviewed Research and Review Articles from Nature Publishing Group},
  pages 187--191. World Scientific, 2011.

\bibitem{wu2012designing}
Hui Wu and Yi~Cui.
\newblock Designing nanostructured si anodes for high energy lithium ion
  batteries.
\newblock {\em Nano today}, 7(5):414--429, 2012.

\bibitem{zhao2019review}
Ying Zhao, Peter Stein, Yang Bai, Mamun Al-Siraj, Yangyiwei Yang, and Bai-Xiang
  Xu.
\newblock A review on modeling of electro-chemo-mechanics in lithium-ion
  batteries.
\newblock {\em Journal of Power Sources}, 413:259--283, 2019.

\bibitem{liu2020computational}
Pengfei Liu, Rong Xu, Yijin Liu, Feng Lin, and Kejie Zhao.
\newblock Computational modeling of heterogeneity of stress, charge, and cyclic
  damage in composite electrodes of li-ion batteries.
\newblock {\em Journal of The Electrochemical Society}, 167(4):040527, 2020.

\bibitem{li2020real}
Dawei Li, Yikai Wang, Bo~Lu, and Junqian Zhang.
\newblock Real-time measurements of electro-mechanical coupled deformation and
  mechanical properties of commercial graphite electrodes.
\newblock {\em Carbon}, 169:258--263, 2020.

\bibitem{heubner2020electrochemical}
C~Heubner, U~Langklotz, C~Laemmel, M~Schneider, and A~Michaelis.
\newblock Electrochemical single-particle measurements of electrode materials
  for li-ion batteries: Possibilities, insights and implications for future
  development.
\newblock {\em Electrochimica Acta}, 330:135160, 2020.

\bibitem{bunjaku2019structural}
Teut{\"e} Bunjaku, Dominik Bauer, and Mathieu Luisier.
\newblock Structural and electronic properties of lithiated si nanowires: An ab
  initio study.
\newblock {\em Physical Review Materials}, 3(10):105402, 2019.

\bibitem{hu2020investigating}
XF~Hu, SJ~Li, J~Wang, ZM~Jiang, and XJ~Yang.
\newblock Investigating size-dependent conductive properties on individual si
  nanowires.
\newblock {\em Nanoscale Research Letters}, 15(1):1--12, 2020.

\bibitem{xu2019heterogeneous}
Rong Xu, Yang Yang, Fei Yin, Pengfei Liu, Peter Cloetens, Yijin Liu, Feng Lin,
  and Kejie Zhao.
\newblock Heterogeneous damage in li-ion batteries: experimental analysis and
  theoretical modeling.
\newblock {\em Journal of the Mechanics and Physics of Solids}, 129:160--183,
  2019.

\bibitem{di2020shuttleworth}
Nicodemo Di~Pasquale and Ruslan~L Davidchack.
\newblock Shuttleworth equation: A molecular simulations perspective.
\newblock {\em The Journal of Chemical Physics}, 153(15):154705, 2020.

\bibitem{cheng2008influence}
Yang-Tse Cheng and Mark~W Verbrugge.
\newblock The influence of surface mechanics on diffusion induced stresses
  within spherical nanoparticles.
\newblock {\em Journal of Applied Physics}, 104(8):083521, 2008.

\bibitem{deshpande2010modeling}
Rutooj Deshpande, Yang-Tse Cheng, and Mark~W Verbrugge.
\newblock Modeling diffusion-induced stress in nanowire electrode structures.
\newblock {\em Journal of Power Sources}, 195(15):5081--5088, 2010.

\bibitem{hao2012diffusion}
Feng Hao, Xiang Gao, and Daining Fang.
\newblock Diffusion-induced stresses of electrode nanomaterials in lithium-ion
  battery: the effects of surface stress.
\newblock {\em Journal of Applied Physics}, 112(10):103507, 2012.

\bibitem{ASJC2019surfacestress}
Amrita Sengupta and Jeevanjyoti Chakraborty.
\newblock Geometry and charging rate sensitively modulate surface
  stress-induced stress relaxation within cylindrical silicon anode particles
  in lithium-ion batteries.
\newblock {\em Acta Mechanica}, 2019.
\newblock in press.

\bibitem{gao2015chemo}
Xiang Gao, Daining Fang, and Jianmin Qu.
\newblock A chemo-mechanics framework for elastic solids with surface stress.
\newblock {\em Proc. R. Soc. A}, 471(2182):20150366, 2015.

\bibitem{dingreville2008interfacial}
R{\'e}mi Dingreville and Jianmin Qu.
\newblock Interfacial excess energy, excess stress and excess strain in elastic
  solids: Planar interfaces.
\newblock {\em Journal of the Mechanics and Physics of Solids},
  56(5):1944--1954, 2008.

\bibitem{schulman2018surface}
Rafael~D Schulman, Miguel Trejo, Thomas Salez, Elie Rapha{\"e}l, and Kari
  Dalnoki-Veress.
\newblock Surface energy of strained amorphous solids.
\newblock {\em Nature communications}, 9(1):1--6, 2018.

\bibitem{jia2020coupling}
Ning Jia, Zhilong Peng, Yazheng Yang, Yin Yao, and Shaohua Chen.
\newblock The coupling effect of surface effect and chemical diffusion in
  lithium-ion battery with spherical nanoparticle electrodes.
\newblock {\em International Journal of Applied Mechanics}, page 2050091, 2020.

\bibitem{sharma2003effect}
P~Sharma, S~Ganti, and N~Bhate.
\newblock Effect of surfaces on the size-dependent elastic state of
  nano-inhomogeneities.
\newblock {\em Applied Physics Letters}, 82(4):535--537, 2003.

\bibitem{cammarata1994surface}
Robert~C Cammarata.
\newblock Surface and interface stress effects in thin films.
\newblock {\em Progress in surface science}, 46(1):1--38, 1994.

\bibitem{lu2018reaction}
Yongjun Lu, Panlong Zhang, Fenghui Wang, Kai Zhang, and Xiang Zhao.
\newblock Reaction-diffusion-stress coupling model for {L}i-ion batteries:
  {T}he role of surface effects on electrochemical performance.
\newblock {\em Electrochimica Acta}, 274:359--369, 2018.

\bibitem{bucci2017effect}
Giovanna Bucci, Tushar Swamy, Sean Bishop, Brian~W Sheldon, Yet-Ming Chiang,
  and W~Craig Carter.
\newblock The effect of stress on battery-electrode capacity.
\newblock {\em Journal of The Electrochemical Society}, 164(4):A645--A654,
  2017.

\bibitem{anand2012cahn}
Lallit Anand.
\newblock A {C}ahn--{H}illiard-type theory for species diffusion coupled with
  large elastic--plastic deformations.
\newblock {\em Journal of the Mechanics and Physics of Solids},
  60(12):1983--2002, 2012.

\bibitem{cui2012finite}
Zhiwei Cui, Feng Gao, and Jianmin Qu.
\newblock A finite deformation stress-dependent chemical potential and its
  applications to lithium ion batteries.
\newblock {\em Journal of the Mechanics and Physics of Solids},
  60(7):1280--1295, 2012.

\bibitem{dora2019stress}
JK~Dora, A~Sengupta, S~Ghosh, N~Yedla, and J~Chakraborty.
\newblock Stress evolution with concentration-dependent compositional expansion
  in a silicon lithium-ion battery anode particle.
\newblock {\em Journal of Solid State Electrochemistry}, 23(8):2331--2342,
  2019.

\bibitem{bower2011finite}
Allan~F Bower, Pradeep~R Guduru, and Vijay~A Sethuraman.
\newblock A finite strain model of stress, diffusion, plastic flow, and
  electrochemical reactions in a lithium-ion half-cell.
\newblock {\em Journal of the Mechanics and Physics of Solids}, 59(4):804--828,
  2011.

\bibitem{di2015diffusion}
Claudio~V Di~Leo, Elisha Rejovitzky, and Lallit Anand.
\newblock Diffusion--deformation theory for amorphous silicon anodes: the role
  of plastic deformation on electrochemical performance.
\newblock {\em International Journal of Solids and Structures}, 67:283--296,
  2015.

\bibitem{liu2011anisotropic}
Xiao~Hua Liu, He~Zheng, Li~Zhong, Shan Huang, Khim Karki, Li~Qiang Zhang, Yang
  Liu, Akihiro Kushima, Wen~Tao Liang, Jiang~Wei Wang, et~al.
\newblock Anisotropic swelling and fracture of silicon nanowires during
  lithiation.
\newblock {\em Nano letters}, 11(8):3312--3318, 2011.

\bibitem{rhodes2010understanding}
Kevin Rhodes, Nancy Dudney, Edgar Lara-Curzio, and Claus Daniel.
\newblock Understanding the degradation of silicon electrodes for lithium-ion
  batteries using acoustic emission.
\newblock {\em Journal of the Electrochemical Society}, 157(12):A1354--A1360,
  2010.

\bibitem{haftbaradaran2011continuum}
Hamed Haftbaradaran, Jun Song, WA~Curtin, and Huajian Gao.
\newblock Continuum and atomistic models of strongly coupled diffusion, stress,
  and solute concentration.
\newblock {\em Journal of Power Sources}, 196(1):361--370, 2011.

\bibitem{bucci2014measurement}
Giovanna Bucci, Siva~PV Nadimpalli, Vijay~A Sethuraman, Allan~F Bower, and
  Pradeep~R Guduru.
\newblock Measurement and modeling of the mechanical and electrochemical
  response of amorphous si thin film electrodes during cyclic lithiation.
\newblock {\em Journal of the Mechanics and Physics of Solids}, 62:276--294,
  2014.

\bibitem{yang2010insertion}
Fuqian Yang.
\newblock Insertion-induced breakage of materials.
\newblock {\em Journal of Applied Physics}, 108(7):073536, 2010.

\bibitem{verbrugge1996modeling}
Mark~W Verbrugge and Brian~J Koch.
\newblock Modeling lithium intercalation of single-fiber carbon
  microelectrodes.
\newblock {\em Journal of The Electrochemical Society}, 143(2):600--608, 1996.

\end{thebibliography}
\bibliographystyle{unsrt}

\clearpage
\begin{appendix}		
\section{Chemical potential for non-ideal mixture}
The non-dimensional flux of the diffusing species in the host material is given by:
\begin{equation}
\tJ_r=-\tilde{D}c\frac{\partial \tmu}{\partial \tr},
\end{equation}
where $\tmu$ is the non-dimensionalized form of the chemical potential of the system ($\mu$). In a coupled chemo-mechanical system, the diffusion being affected by the mechanical stresses, $\mu$ becomes a function of both concentration and mechanical stresses ($\mu=\mu_0+\mu_{\rm s}$, where $\mu_0$ is the stress-indepdendent part of the chemical potential and $\mu_{\rm s}$ is the stress-dependent part). Keeping aside the stress-enhanced diffusion for the moment, we concentrate on the different forms of $\mu_0$ considered by different researchers, and their impact on the present problem.\\

Chemical potential is the change in free energy of the system due to an infinitesimal change in the concentration of the diffusing species, mathematically described in Eq. \eqref{mu}. \\
For an ideal mixture,
\begin{equation} \label{Gideal}
\Delta G^{\rm ideal}_{\rm mix}=R_{\rm g}T[x_A {\rm ln}x_A+x_B {\rm ln}x_B],
\end{equation}
where $x_A$ and $x_B$ are the mole fractions of the two components in the mixture. \\
For a non-ideal/real mixture,
\begin{equation} \label{Greal}
\Delta G^{\rm real}_{\rm mix}=R_{\rm g}T[x_A {\rm ln}x_A+x_B {\rm ln}x_B+\kappa x_A x_B].
\end{equation}
The extra term in Eq. \eqref{Greal} is due to the deviation from ideality, and $\kappa$ is the dimensionless interaction parameter.\\
In the present case, 
\begin{equation}
\Delta G=R_{\rm g}T[c {\rm ln}c+(1-c){\rm ln}(1-c)+\kappa c(1-c)],
\end{equation}
which when partially differentiated with respect to $c$ gives the chemical potential as:
\begin{equation} \label{mulu}
\mu_0=R_{\rm g}T[\kappa(1-2c)+{\rm ln}\frac{c}{1-c}].
\end{equation}
This form of $\mu_0$ has been used by Lu et. al. \cite{lu2018reaction}.\\

Haftabaradaran et. al. \cite{haftbaradaran2011continuum} and Cui et, al. \cite{cui2012finite} studied solutions with high solute concentrations. In the continuum modelling of diffusion in solids, the stress-independent part of the chemical potential is expressed as \cite{haftbaradaran2011continuum}
\begin{equation}
\mu_0=\mu_{\rm s}-\mu_{\rm v},
\end{equation}
where $\mu_i=\mu_i^0+R_{\rm g}T {\rm ln}a_i$ ($i={\rm s,v}$). Here $\mu^0_{\rm v}$ is the free energy of the host solid having 1 mole of vacant interstitial sites, and $\mu^0_{\rm s}$ is the free energyof the same system when the interstitial sites are occupied by the diffusing atoms; $a_i$ is the interaction energy, and is expressed as:
\begin{equation}
a_{\rm s}=\gamma_{\rm s}c \qquad a_{\rm v}=\gamma_{\rm v}(1-c),
\end{equation}
where $\gamma_{\rm s}$ and $\gamma_{\rm v}$ are the activity coefficients of the solute atoms and vacancies, respectively. Therefore, chemical potential becomes
\begin{equation} \label{muhafta}
\mu_0=\mu^0_{\rm sv}+R_{\rm g}T {\rm ln} \frac{\gamma_{\rm s}c}{\gamma_{\rm v}(1-c)},
\end{equation}
where $\mu^0_{\rm sv}=\mu^0_{\rm s}-\mu^0_{\rm v}$ is a constant. Equation \eqref{muhafta} can be further simplifies to give
\begin{equation} \label{mucui}
\mu_0=\mu^0_0+R_{\rm g}T {\rm ln}(\gamma c),
\end{equation}
where $\displaystyle \gamma=\frac{1}{1-c} {\rm exp}\left( \frac{\Delta \tau_0}{R_{\rm g} T}\right)$ such that, $\Delta \tau_0= 2(A_0-2B_0)c-3(A_0-B_0)c^2$ \cite{cui2012finite} and $\mu^0_0=\mu^0_{\rm sv}$. Cui et. al. \cite{cui2012finite} used this form of stress-independent chemical potential and worked on the stress-dependent $\mu$-form.\\
\begin{figure} [h!]
	\centering
	\begin{subfigure}[b]{0.5\textwidth}
		\includegraphics[width=\textwidth]{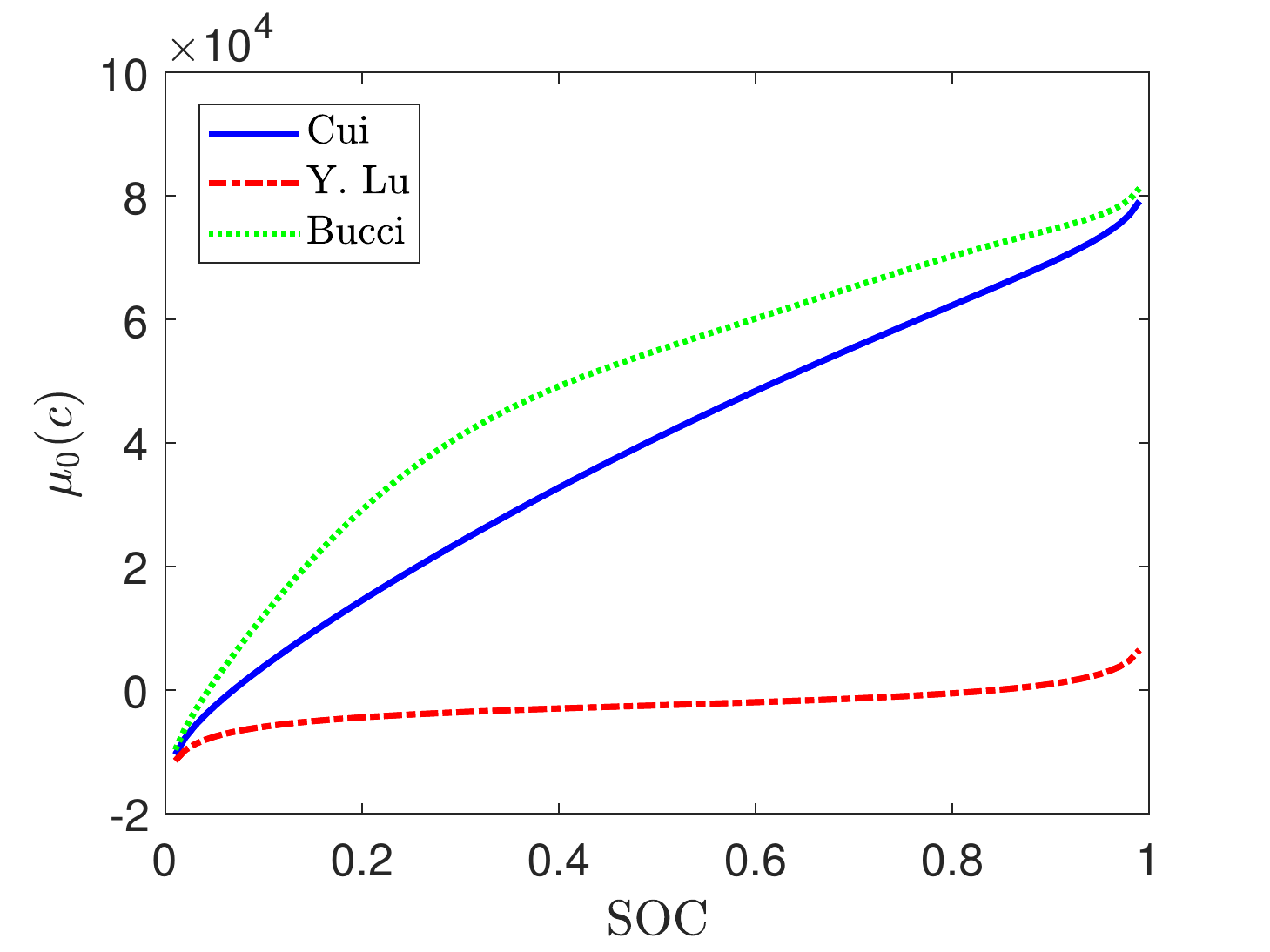}
		\caption{}
	\end{subfigure}%
	\begin{subfigure}[b]{0.5\textwidth}
		\includegraphics[width=\textwidth]{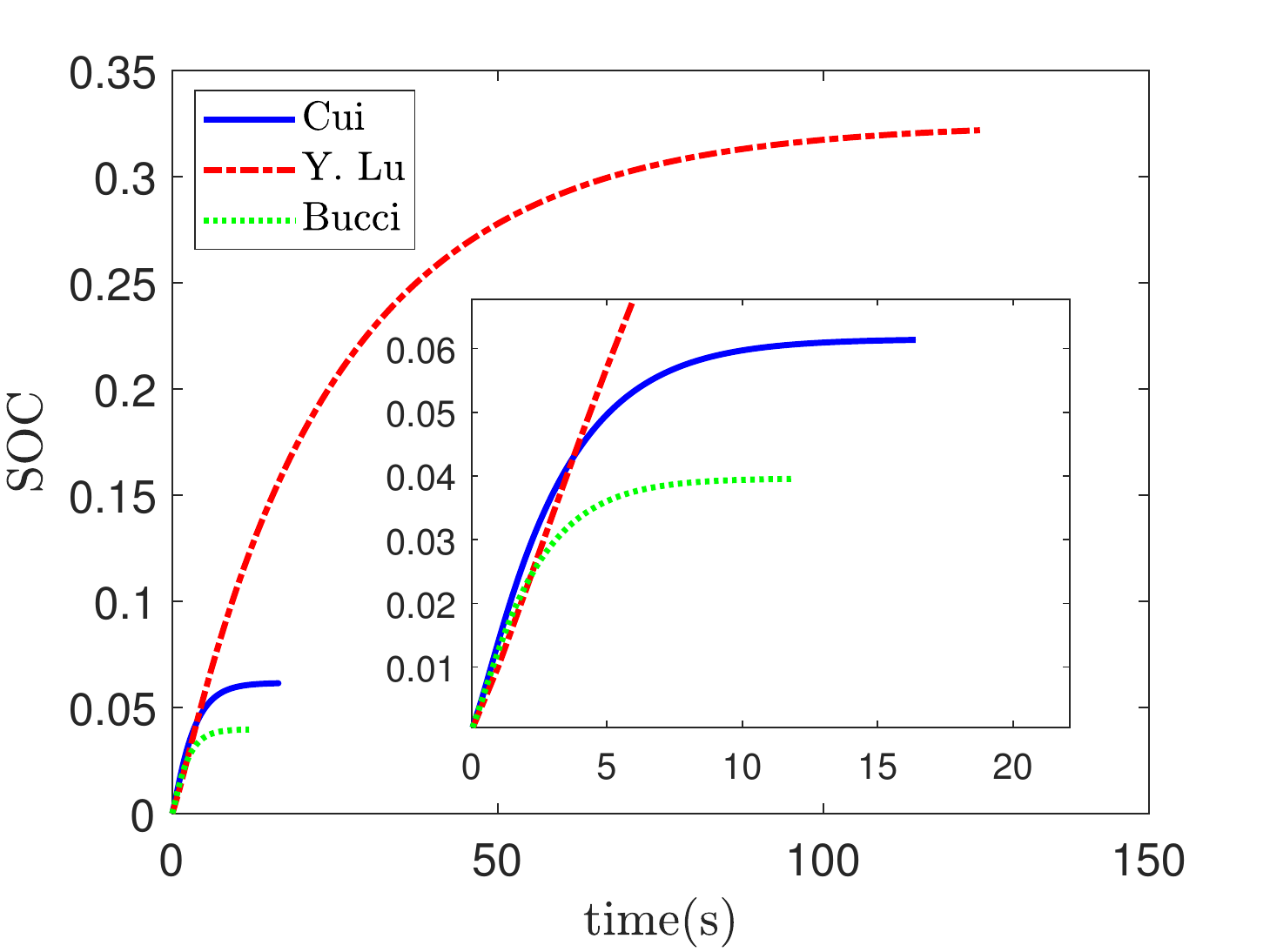}
		\caption{}
	\end{subfigure}
	\caption{(a) Variation of stress-independent part of chemical potential with SOC for different forms of $\mu_0(c)$ as considered by Cui et. al.  \cite{cui2012finite}, Lu et. al.  \cite{lu2018reaction}, and Bucci et. al.  \cite{bucci2014measurement}; (b) SOC achieved by a particle of size 50 nm, in the potentiostatic charging condition, considering different forms of $\mu_0(c)$. } \label{chempotcomparison}
\end{figure}
Bucci et. al. \cite{bucci2014measurement} and Leo et. al. \cite{di2015diffusion} on the other hand, followed Verbrugge and Koch's model \cite{verbrugge1996modeling} and defined $\mu$ as 
\begin{equation} \label{mubucci}
R_{\rm g}T {\rm ln}\gamma= \sum_{{\sf n}=2}^{\sf N} \Omega_{\sf n}{\sf n}c^{\sf n-1},
\end{equation}
where $\Omega_{\sf n}$ are a set of coefficients, determined experimentally. \\

In Fig. \ref{chempotcomparison} (a) we plot the different forms of stress-independent part of chemical potential against SOC. While the evolution of $\mu_{0 {\rm Cui}}$ and $\mu_{0 {\rm Bucci}}$ are steep and comparable, the variation of $\mu_{0 {\rm Y.Lu}}$ with the growing concentration of Li is very slow. These differnet forms of $\mu_0$ provide satisfactory results in case of galvanostatic charging/discharging condition. But, in case of potentiostatic charging, it is observed that the equilibrium SOC reached is directly affected by the form of $\mu_0$ considered. As illustrated in Fig. \ref{chempotcomparison} (b), when we use $\mu_{0 {\rm Cui}}$ and $\mu_{0 {\rm Bucci}}$, we achieve a very low value of equilibrium SOC (below 6 $\%$). On the contrary, using $\mu_{0 {\rm Y.Lu}}$ we achieve an equilibrium SOC upto 33 $\%$. Hence, in the present study, for better representation of results, we consider the form of $\mu_0$ as taken by Lu et. al. \cite{lu2018reaction}.

\end{appendix}

\end{document}